\documentstyle[prb,aps,epsf,floats]{revtex}
\newcommand{\beq}{\begin{equation}}
\newcommand{\eeq}{\end{equation}}
\newcommand{\beqr}{\begin{eqnarray}}
\newcommand{\eeqr}{\end{eqnarray}}
\newcommand{\e}{{\epsilon}}

\newcommand{\w}{{\omega}}

\def\vtau{{\vec\tau}}
\def\hi{{\hat i}}
\def\hj{{\hat j}}

\def\bp{{\mathbf p}}
\def\bI{{\mathbf I}}

\def\bM{{\mathbf M}}
\def\bq{{\mathbf q}}
\def\br{{\mathbf r}}

\def\bG{{\mathbf G}}
\def\bGamma{{\mathbf \Gamma}}

\def\bk{{\mathbf k}}

\newcommand{\sigmab}{\mbox{\boldmath $\sigma $}}

\newcommand{\p}{{\partial}}

\def\half{{1\over2}}

\def\ua{\uparrow}
\def\da{\downarrow}
\def\eqa{\begin{eqnarray}}
\def\eea{\end{eqnarray}}
\def\cE{{\cal E}}
\def\cP{{\cal P}}
\def\cT{{\cal T}}
\def\cL{{\cal L}}
\def\cN{{\cal N}}
\def\ve{{\varepsilon}}
\def\a{{\alpha}}
\def\b{{\beta}}
\def\g{{\gamma}}
\def\d{{\delta}}
\def\t{{\theta}}
\def\tu{{\tilde u}}
\def\eurlet{Europhys. Lett.}
\begin{document}
\draft \flushbottom \twocolumn[
\hsize\textwidth\columnwidth\hsize\csname
@twocolumnfalse\endcsname
\title{ A Solvable Regime of Disorder and Interactions in Ballistic Nanostructures,
Part I: Consequences for Coulomb Blockade}
\author{Ganpathy Murthy$^1$, R. Shankar$^2$, Damir Herman$^3$, and Harsh Mathur$^3$}
\address{$^1$Department of Physics and Astronomy, University of Kentucky,
Lexington KY 40506-0055\\ $^2$ Department of Physics, Yale
University, New Haven CT 06520 \\ $^3$Physics Department, Case
Western Reserve University, Cleveland, OH 44106-7079}
\date{\today}
\maketitle
\begin{abstract}
We provide a framework for analyzing the problem of interacting
electrons in a ballistic quantum dot with chaotic boundary conditions
within an energy $E_T$ (the Thouless energy) of the Fermi
energy. Within this window we show that the interactions can be
characterized by Landau Fermi liquid parameters. When $g$, the
dimensionless conductance of the dot, is large, we find that the
disordered interacting problem can be solved in a saddle-point
approximation which becomes exact as $g\to\infty$ (as in a large-$N$
theory).  The infinite $g$ theory has two phases as a function of the
Landau parameter $u_m$ in a channel with angular momentum $m$: A
weak-coupling phase where constant charging and exchange interactions
dominate the low-energy physics, as in previous ``Universal
Hamiltonian'' treatments, and a strong-coupling phase characterized by
the same order parameter as in the Pomeranchuk transition in clean
systems (a spontaneous interaction-induced Fermi surface distortion),
but smeared and pinned by disorder.  Thus, both interactions and
disorder are crucial to the existence of these phases. At finite $g$,
the two phases and critical point evolve into three regimes in the
$u_m-1/g$ plane -- weak- and strong-coupling regimes separated by
crossover lines from a quantum-critical regime controlled by the
quantum critical point.  In this, the first of a two part series, we
focus on the consequences of this picture for Coulomb Blockade
experiments. We employ analytical and numerical methods to predict the
statistics of single-particle levels, Coulomb Blockade peak spacings,
conductance peak heights and quasiparticle widths. We show that in the
strong-coupling and quantum-critical regions, the quasiparticle
acquires a width of the same order as the level spacing $\Delta$
within a few $\Delta$'s of the Fermi energy due to coupling to
collective excitations. In the strong coupling regime if $m$ is odd,
the dot will (if isolated) cross over from the orthogonal to unitary
ensemble for an exponentially small external flux, or will (if
strongly coupled to leads) break time-reversal symmetry
spontaneously. For any $m$, the peak spacing distribution becomes
broader than expected in previous works and even has support at
negative values, which in turn is correlated with small peak
heights. Ballistic/chaotic quantum dots afford us unrivalled
theoretical and experimental control over the problem of simultaneous
disorder and interactions due to the $1/g$ expansion and our ability
to vary disorder and interaction much more readily than in the bulk.
\end{abstract}
\vskip 1cm ]

\section{Introduction}
\label{intro}

The problem of treating electronic interactions in mesoscopic
systems brings together two very interesting subfields of
condensed matter physics. On the one hand, bulk systems with
interactions and disorder can show unexpected
phenomena\cite{2dmit}, and are the subject of ongoing and vigorous
investigation. On the other hand, mesoscopic
systems\cite{mesoscopics-review} can show behavior which is not
present in bulk systems. Two examples are the oscillations as a
function of gate voltage $V_g$ in the tunnelling conductance of a
quantum dot (QD) weakly coupled to the leads (the Coulomb Blockade (CB)
regime\cite{CB,qd-reviews,small-rs-expt,large-rs-expt,others-expt}),
and persistent currents\cite{persist-expt} in small metallic rings
subject to a Aharanov-Bohm flux.

In recent months we have developed a formalism that can tackle the
problem of mesoscopic, disordered, strongly correlated
systems\cite{qd-us1,qd-us2}.  We showed that there exists a
controlled and complete solution to this  problem provided the
quantum dot (or ring) is ballistic, meaning that the disorder
comes from chaotic collisions with the walls of the mesoscopic
structure, and not from impurities inside it. The solution becomes
exact in the limit when the dimensionless conductance $g$ of the
quantum dot becomes large\cite{qd-us1,qd-us2}. In this paper we
focus on the details of our approach in CB regime, while Part
II\cite{partII} of this paper will analyze the effects on
persistent currents (a short report of which has
appeared\cite{short}).  Experimental Coulomb Blockade
samples\cite{small-rs-expt,large-rs-expt,others-expt} are
ballistic and do have fairly large conductances ($g\approx 5-20$).
Our predictions are directly applicable to such samples even  if
the interactions between electrons are strong, which can be
achieved by lowering  the electron density below that of  present
samples.

Let us first look at some important length, time, and energy scales to
set up the problem. We will confine our analysis to two dimensions,
since experimental samples are made by confining a two-dimensional
electron gas (2DEG) laterally by means of gates.  We will focus
attention on ballistic QD's for which the bulk mean free path $l$ is
much larger than system size $L$. Ignoring the (weak) coupling to the
leads produces sharp single-particle levels in the QD whose mean
spacing is $\Delta$. A famous conjecture by Bohigas, Giannoni, and
Schmidt\cite{bgs} states that all levels within a Thouless energy
$E_T=\hbar v_F/L$ of each other have correlations controlled by Random
Matrix Theory (RMT)\cite{RMT}.  The analogous conjecture has been
proved for the case of diffusive QD's\cite{alt1}, and there are
ongoing efforts to prove it for the ballistic
case\cite{efetov-ballistic}. We will be interested in applying RMT to
states that lie within $E_T$ of the Fermi energy.  Of particular
interest to us are two RMT ensembles\cite{RMT}, the Gaussian
Orthogonal Ensemble (GOE) which applies to time-reversal invariant
systems, and the Gaussian Unitary Ensemble (GUE) which applies to
systems in which time-reversal is broken by an external magnetic
flux. We will assume that spin-orbit coupling is negligible throughout
this work. Also, because of a variety of factors, the coupling of the
external magnetic field to the spin is small compared to the orbital
effects in $GaAs$, the standard material used to make quantum dots. To
simplify matters, we will set the spin-$B$-field coupling to zero in
what follows. The dimensionless conductance $g$ (also called the
Thouless number) is defined by the ratio $g=E_T/\Delta$. Two energies
inherited from the bulk 2DEG are the bandwidth $E_b$ and the Fermi
energy $E_F$. The strength of interactions in the bulk 2DEG is
characterized by the dimensionless number $r_s=a/a_0$, where
$a=1/\sqrt{\pi \rho}$ is the typical distance between neighboring
particles ($\rho$ is the density) while $a_0=\hbar^2/me^2$ is the Bohr
radius. The band effective mass must be used in computing $a_0$ as
must the dielectric constant. Since the system is finite, capacitive
effects produce a charging energy $U_0=e^2/C$, where $C$ is the
capacitance.

The experiments that contributed greatly to our understanding have
been on the  the zero-bias conductance of a QD weakly coupled to
the leads as a function of $V_g$ at $T=0$. At a generic  value of
$V_g$ the ground state has a definite number of particles $N$ and
energy $\cE_N$. If the chemical potential
$\mu=\cE_{N+1}-\cE_N=\alpha eV_g$ ($\alpha$ is a
geometry-dependent ``leverage'' factor\cite{qd-reviews}) the free
energies of the $N$ and $N+1$-particle states are degenerate, and
a tunneling peak occurs at zero bias. Successive peaks are
separated by the second difference of $\cE_N$, called $\Delta_2$,
the distribution of which is
measured\cite{small-rs-expt,large-rs-expt}.

To analyze the problem theoretically, it is simplest to consider a
theory within the Thouless shell around $E_F$ (defined by
$|\ve-E_F|\le E_T/2$). In this shell all the statistical
properties of the single-particle energies and wavefunctions can
be obtained from RMT. The generic Hamiltonian in this shell can be
written as
\begin{eqnarray} H&=&\sum_{\alpha} c^{\dag}_{\alpha}c_{\alpha}
\varepsilon_{\alpha}+{1 \over 2}\sum_{\alpha \beta \gamma
\delta}V_{\alpha \beta \gamma
\delta}c^{\dag}_{\alpha}c^{\dag}_{\beta}c_{\gamma}c_{\delta}\label{hran}
\end{eqnarray}
where $\varepsilon_{\alpha}$ are single-particle levels that obey
RMT statistics, have a mean spacing $\Delta$ and range from
$-g\Delta /2$ to $g\Delta /2$, and $V_{\a\b\g\d}$ represents a
two-body interaction. In the following we will supress spin for
simplicity, pointing out how its restoration modifies various
results. Different forms of $V_{\a\b\g\d}$ turn out to represent
very different physics.

The simplest model for interactions in a QD has a constant
charging energy\cite{CB,qd-reviews} $U_0$: \beq
H_U=\sum\limits_{\a,s} \ve_\a c^{\dagger}_{\a,s} c_{\a,s}
+{U_0\over2}{\hat N}^2 \eeq which corresponds to the choice
$V_{\alpha \beta \beta \alpha}=U_0\equiv
 u_0\ \Delta $,
(independent of $\a$ and $\b$ with  all other couplings not of
this form vanishing).
 This model   predicts a bimodal
distribution for $\Delta_2$: Adding an electron above a
doubly-filled (spin-degenerate) level costs $U_0+\ve$, with $\ve$
being the energy to the next single-particle level. Adding it to a
singly occupied level costs $U_0$. While the second contribution
gives a Dirac delta-function peak at $U_0$, the first contribution
is the distribution of nearest neighbor level separation $\ve$,
which is known\cite{RMT} from RMT to have a width of the order of
the mean single-particle level spacing $\Delta$. The puzzle that
stimulated recent theoretical developments in the field is that
numerics\cite{exact,hf} and
experiments\cite{small-rs-expt,large-rs-expt} produce
distributions for $\Delta_2$ which do not show any bimodality, and
are much broader.

 One interesting option is to consider   $V_{\alpha \beta \gamma \delta}$ as independent gaussian
variables\cite{2brim}. Our work is more closely related to
another,  the Universal Hamiltonian\cite{H_U,univ-ham}, wherein (
for the spinful case) a term coupling to total ${\vec S}^2$ and a
Cooper coupling also appear: \beqr H_U=&\sum\limits_{\a,s} \ve_\a
c^{\dagger}_{\a,s} c_{\a,s} +{U_0\over2}{\hat N}^2-{J\over2}{\vec
S}^2 \nonumber\\ &+\lambda\bigg(\sum\limits_{\a}
c^{\dagger}_{\a,\ua}c^{\dagger}_{\a,\da}\bigg)\bigg(\sum\limits_{\b}c_{\b,\da}c_{\b,\ua}\bigg)
\label{hu} \eeqr The rationale for the above choice of
interactions is the following: As we will see below, only the
``diagonal'' matrix elements in which the indices are pairwise equal
survive the ensemble average and the variances of all the matrix
elements are small \beq <V_{\a\b\g\d}^2>-<V_{\a\b\g\d}>^2\simeq
{\Delta^2\over g^2} \eeq where the indices now describe orbital and
spin variables. Thus for large $g$, the interaction may be
approximated by its ensemble average, which is just the Universal
Hamiltonian. Apart from this motivation, calculations based on the
Universal Hamiltonian\cite{H_U,univ-ham} have proven very successful
at describing experiments\cite{small-rs-expt} on quantum dots with
small $r_s\approx 1$ at a quantitative level, once experimental noise
is subtracted out and finite-temperature
effects\cite{usaj-finite-T,pk-ht-T-exchange,yoram} are taken into
account.

Our work deals with situations where $r_s$ is a lot larger than in
current samples, in which case other unexpected phenomena appear
possible.  Our choice of $V_{\a\b\g\d}$ follows from the assumption of
Landau Fermi liquid interactions at the Thouless
energy\cite{qd-us1,qd-us2}, which we justify using renormalization
group arguments. The Landau interactions\cite{landau,agd} are
parametrized by couplings $u_m$ in every angular momentum channel
$m$. For the spinful case the set of parameters is doubled, with one
for the charge and one for the spin. Keeping only $u_0^{charge}$ and
$u_0^{spin}$ yields the Universal Hamiltonian (without the Cooper
term).  The retention of all the Landau parameters leads to the
emergence of strong-coupling regimes in $m\ne0$ angular momentum
channels\cite{qd-us1,qd-us2}, which exhibit a spontaneous deformation
of the Fermi surface (suitably smeared out and pinned by disorder) as
well as the possibility of time-reversal violation. The transition was
originally discovered in the clean bulk limit by
Pomeranchuk\cite{pomeranchuk}. While these are true second-order
quantum phase transitions in the limit $g\to\infty$, they are replaced
by sharp crossovers for finite but large $g$.  This paper presents in
detail the physics of these new phases, the transitions/crossovers
between the weak- and strong-coupling regime via the quantum-critical
regime and their attendant signatures. We concentrate on
charge-channel instabilities in the spinful system; spin-channel
instabilities necessarily involve the exchange coupling $J$ and will
be the subject of future work. The purpose of this set of two papers
is to supply all the details left out of the previous brief
presentations\cite{qd-us1,qd-us2,short}, as well as to present
numerical corroboration of the earlier analytical results, many new
results, and experimental signatures.

The plan of the paper is as follows.  In section II we present the
renormalization group (RG) argument\cite{rg-shankar} for starting
with Landau Fermi liquid interactions in the Thouless band, with
all the assumptions and caveats. In particular, we will show why
these assumptions are reasonable for ballistic QD's but might fail
for QD's deep in the diffusive limit\cite{altshuler-aronov}. In
Section III we present a further RG analysis inside the Thouless
shell (modeled after Ref. \cite{rg-us}), this time integrating out
exact eigenstates of the disordered single-particle Hamiltonian.
To leading order in $1/g$ we will see that the RG flow to one-loop
order implies the phase transitions mentioned above. However, some
questions about the effects of higher loops are left open, as is
the very nature of the  strong-coupling phase. These issues are
resolved in Section IV, where we show that the problem can be
mapped on to a large-$N$ theory, with $g$ playing the role of $N$.
\footnote{This large-$N$ nature has two manifestations, one which
is best expressed by saying that a certain class of diagrams
dominate over others, and another in which a large number (here
$g^2$) appears in front of the effective action, showing that the
saddle-point approximation is a good one.} The largeness of $g$
allows us to fully control the calculation at at strong-coupling
and not only confirm the phase transition discovered in
perturbative RG\cite{qd-us1} but also to probe the strong coupling
phase in considerable detail\cite{qd-us2}. In the next and longest
section (Section V) we explore the physical consequences of the
large-$g$ solution: The nature of the order parameter, the
physical signatures of the strong-coupling regime and the
crossover quantum-critical regime\cite{critical-fan}, and some
curious properties of the weak-coupling regime.  Numerical methods
which are capable of accessing all regions of the phase diagram
(except the quantum critical regime) corroborate
 the analytical methods and also complement them, for example in
 the computation of
 level statistics and the CB peak-spacing statistics in
the strong-coupling regime.

We predict that if the system is in the strong-coupling regime of
an even $m$ channel it will display a broadening of the CB peak
spacing distribution, correlations between small peak-heights and
small peak-spacings, and a diamagnetic persistent current (which
is not expected theoretically\cite{persist-int-th,persist-puzzle}
in non-superconducting materials, but is seen
experimentally\cite{persist-expt}).

In the strong-coupling regime of an odd $m$ channel, in addition
to all the above effects (but with a paramagnetic persistent
current), the system can be mapped on to the
Caldeira-Leggett\cite{CL} model of a particle in a double-well
potential subject to ohmic dissipation. The solution to this
model\cite{chakravarty,bray-moore} shows that the $m$ odd system
can be driven through a quantum phase transition which
spontaneously breaks time-reversal symmetry by changing the
coupling of the quantum dot to the leads.  (This really is a phase
transition since the quantum coherence of the dot with the
infinite reservoir renders a sharp transition possible).

In this section we will also show that while an effective
single-particle description works in the weak-coupling regime, it
fails in the quantum-critical and strong-coupling regimes even at
low energies of order a few $\Delta$. In these latter regions we
will make a connection to ideas of ``Fock-space
localization/delocalization''\cite{fock-loc,fock-loc2}, which
captures the crossover of the spectral width of the quasiparticle
to a Breit-Wigner form as the quasiparticle energy is increased.
This  long section ends with its own  summary. In Section VI our
conclusions are presented, as are  discussions of  how our
approach might be useful in other problems involving disorder and
interactions. Appendices  A and B contain some additional details.

\section{Effective Interactions at the Thouless Energy}

In this section we will give a brief introduction to Fermi liquid
theory\cite{landau,agd} in the bulk, describe its
instabilities\cite{pomeranchuk}, and see how it must be modified
to account for disorder. We will argue that a hybrid Hamiltonian,
with a single-particle part encoding the chaotic nature of the
single-particle states, and a part with Fermi liquid interactions,
is the most natural starting point for analyzing ballistic
mesoscopic structures in the Thouless band.

\subsection{The Clean Limit in the Bulk}
\label{clean-bulk}

It is well-known that in a clean two-dimensional bulk system, no
matter what interactions one starts with in the bare Hamiltonian,
provided no superconducting or density wave instabilities
intervene, the effective interactions are of the Landau Fermi
liquid form\cite{landau,agd,rg-shankar} in a sufficiently thin
shell near the Fermi surface defined by $|\bk|-k_F\le k_{max}$.
This corresponds to an energy cutoff $\Lambda=v_F k_{max}$. The
entire Hamiltonian in such a thin shell can be written for the
spinless case as

\beqr H_{FL}=&\sum\limits_{\bk} \ve_0(\bk) c^{\dagger}(\bk)c(\bk)
+{1\over 2N_0}\sum\limits_{\bk,\bk',\bq} u(\t-\t') \nonumber\\
&\times :c^{\dagger}(\bk-\bq)c(\bk)c^{\dagger}(\bk'+\bq)c(\bk'):
\label{hfl}\eeqr where $\ve_0(\bk)$ is the renormalized
quasiparticle energy measured from the Fermi surface, $c(\bk),\
c^{\dagger}(\bk)$ are canonical anticommuting fermion operators,
$N_0$ is the single-particle density of states, the $::$ sign
stands for normal-ordering (subtracting the average in the ground
state), $\t,\ \t'$ are the angles of the two-dimensional vectors
$\bk,\ \bk'$, and $u(\t-\t')$ is the (dimensionless) Fermi liquid
interaction function. It is understood that $\bk,\ \bk'$ are to be
summed only over states in the shell, and that $|\bq|\le k_{max}$.
The main features of the Fermi liquid interaction are that
$u(\t-\t')$ does not depend on the radial magnitudes of $\bk,\
\bk'$ but only on the angles, and that the interaction shows
mostly forward scattering. Rotational invariance has been used to
express $u$ as a function of the difference of the angles. One
then Fourier decomposes $u$ to obtain

\beq u(\t-\t')=u_0+\sum\limits_{m=1}^{\infty} u_m \cos{m(\t-\t')}
\eeq where the $u_m$ are known as Landau parameters. The Fermi
liquid interaction function also determines the energy to add a
quasiparticle with momentum $\bk$ in the presence of a background
of excited quasiparticles/quasiholes specified by $\d n(\bk')$
(which is the deviation in occupation from the ground state)

\beq \ve(\bk)=\ve_0(\bk)+{1\over N_0}\sum\limits_{\bk'}
u(\t-\t')\d n(\bk') \eeq and the Landau total energy functional

\beq \cE(\{\d n\})=\sum\limits_\bk \ve_0(\bk)\d n(\bk) +{1\over
2N_0}\sum\limits_{\bk\bk'} u(\t-\t') \d n(\bk)\d n(\bk')
\label{spinless-E-functional} \eeq

Landau's original derivation\cite{landau,agd} rests on phase space and
adiabatic continuity arguments and also predicts a decay rate for
quasiparticles of order $\ve^2/E_F$, which has been neglected
above. The same result can also be derived by integrating out high
energy states (single-particle states far from the Fermi surface) in a
renormalization group (RG) approach, as was shown by one of us a
decade ago\cite{rg-shankar}. In this approach, at a given stage of the
RG one has a theory with a certain cutoff $k_{max}$.  One then
integrates out a thin ``shell'' of momenta or width $\delta k_{max}$,
thereby obtaining an effective theory with a cutoff $k_{max}-\delta
k_{max}$. One demands that the new theory give the same answer for all
physical Green's functions in the low-energy sector as the theory with
cutoff $k_{max}$. To achieve this the coupling constants must flow as
one changes the cutoff.  If a certain coupling constant increases as
high-energy states are integrated out, it will end up dominating the
low-energy physics no matter how small it was initially. Such a
coupling is called ``relevant''. In the opposite case a coupling may
shrink as high-energy states are integrated out, in which case it is
unimportant for low-energy physics, and is called ``irrelevant''.
Couplings which do not flow are called ``marginal''. It turns out that
all the Landau parameters are marginal for a clean system in the bulk,
whereas all other types of couplings, with one exception are
irrelevant. The exception is the coupling in the BCS channel, i.e.,
between particle of opposite momenta. These are irrelevant if
repulsive and relevant if attractive. Since no superconductivity has
ever been detected in the $GaAs$ 2DEG's which are the basis of the
ballistic quantum dots we focus on, we will assume that the BCS
instability is absent.

Thus, the Landau theory defines a fixed point (in fact a whole
class of fixed points) for the clean electron gas.

\subsection{Fermi Liquid Parameters for the Spinful Case}
\label{flt-spin}

In the spinful case the Fermi liquid interaction function is to be
thought of as a matrix in spin space\cite{agd}. Consider the
energy to add a particle in a momentum state $\bk$ with a density
matrix $\rho_{ss'}(\bk)$. This energy is

\beq
\ve_{\{\rho\}}(\bk)=Tr(\ve(\bk)\rho(\bk))=\sum\limits_{ss'}\ve_{ss'}\rho_{s's}
\eeq

The energy matrix $\ve_{ss'}(\bk)$ depends on the occupations and
spin states of other quasiparticles, or in other words, the
density matrices $\delta n_{s_1s_1'}(\bk')$ in the following way

\beqr \ve_{ss'}(\bk)=&\ve_{0,ss'}(\bk)+\nonumber\\ &{1\over
N_0}\sum\limits_{s_1s_1',\bk\bk'} u_{ss',s_1s_1'}(\t-\t')\delta
n_{s_1s_1'}(\bk') \label{fli-spin1}\eeqr

with the corresponding total energy functional

\beqr &\cE(\{\d n\})=\sum\limits_{ss'\bk}\ve_{0,ss'}(\bk)\d
n_{s's}(\bk)\nonumber\\ &+{1\over
2N_0}\sum\limits_{ss's_1s_1',\bk\bk'} u_{ss',s_1s_1'}(\t-\t')\d
n_{ss'}(\bk)\d n_{s_1s_1'}(\bk') \label{spinful-E-functional}\eeqr

We will restrict ourselves to the case with
spin-rotation-invariance, which leads to a restricted form for the
interaction function\cite{agd}

\beq
u_{ss',s_1s_1'}(\t-\t')=\Phi(\t-\t')\delta_{ss'}\delta_{s_1s_1'}+Z(\t-\t')\vtau_{ss'}\cdot\vtau_{s_1s_1'}
\label{fli-spin2}\eeq

where the $\vtau=(\tau^{(x)},\tau^{(y)},\tau^{(z)})$ are the Pauli
spin matrices, and the $\Phi$ and $Z$ are the Fermi liquid
interaction functions in the charge and spin channels
respectively. As usual, these can be Fourier expanded to obtain
the charge and spin channel Landau parameters

\beqr
\Phi(\t-\t')=&\Phi_0+\sum\limits_{m=1}^{\infty}\Phi_m\cos{m(\t-\t')}\\
Z(\t-\t')=&Z_0+\sum\limits_{m=1}^{\infty}Z_m\cos{m(\t-\t')}
\label{fli-spin3} \eeqr

The $\Phi_m$ and $Z_m$  are identical to the parameters used by
Pines and Nozieres\cite{nozieres}, \beqr F^{s}_m=&\Phi_m\\
F^{a}_m=&Z_m \eeqr

The Landau parameters depend on the relative strength of the
interaction and kinetic energies, commonly characterized by the
dimensionless number $r_s=a/a_0$, where $a=1/\sqrt{\pi \rho}$ is
the typical distance between neighboring particles ($\rho$ is the
density) while $a_0=\hbar^2/me^2$ is the Bohr radius. The band
effective mass must be used in computing $a_0$ as must the
dielectric constant. The dependence of the Landau parameters on
$r_s$ has been the subject of investigation by quantum Monte Carlo
methods\cite{kwon-ceperley}. For the largest value investigated
$r_s=5$, the results in our normalization (differing from that of
ref. \cite{kwon-ceperley} by a factor of two) are \beqr
\Phi_0=&-1.85\\ Z_0=&-0.25\\ \Phi_1=&0.06\\ Z_1=&-0.135\\
\Phi_2=&-0.25\label{fli-parameters}\\ Z_2=&0.16 \eeqr

\subsection{Instabilities of the Clean Fermi Liquid}
\label{fl-instabilities}

It is well-known that the Fermi liquid is unstable towards the
introduction of an attractive coupling in the Cooper channel which
leads to a gapped superconducting ground state. The Fermi liquid also
has other instabilities for certain values of the Fermi liquid
parameters, a fact first pointed out by
Pomeranchuk\cite{pomeranchuk}. There has been a revival of interest in
the Pomeranchuk transition in clean bulk 2D systems
recently\cite{varma,oganesyan}.

Let us first consider the spinless case. We consider a deformation
of the Fermi surface by an amount $r(\t)$ in the direction $\t$.
To be precise, we consider the following deformation

\beqr \d n(\bk)=\left\{ \begin{array}{lll}
                    1,& 0\le\ve_0(\bk)\le r(\t),&r\ge 0\\
                    -1,&r(\t)\le\ve_0(\bk)\le 0,&r\le 0
                    \end{array}
             \right.
\eeqr

Using the replacement $\sum_\bk \to N_0\int d\ve$ we can calculate the
energy cost for this deformation using the Landau energy
functional, Eq. (\ref{spinless-E-functional}), \beq {\cE\over
N_0}=\int\limits_{0}^{2\pi} {d\t\over2\pi} {r(\t)^2\over2}
+\half\int\limits_{0}^{2\pi} {d\t d\t'\over(2\pi)^2}
u(\t-\t')r(\t)r(\t') \eeq

To isolate a particular Fermi liquid channel of angular momentum
$m\ne0$ we choose $r(\t)=r_0\cos{m\t}$ to obtain the total energy
for $m\ne0$ \beq {\cE\over N_0}={r_0^2\over 4}(1+u_m/2)
\label{pom-spinless}\eeq

It is clear that for $u_m\le-2$ the energy is negative, indicating
an instability of the undeformed Fermi liquid. This is the
Pomeranchuk instability\cite{pomeranchuk}.

In the spinful case the ``deformation'' can be in the charge or
the spin channel. Let us consider the instability in the spin
channel for illustration. Now $\d n$ must be a density matrix in
spin space, and we can choose, for example,

\beqr \d n(\bk)=\left\{ \begin{array}{lll}
                    \tau^{z},& 0\le\ve_0(\bk)\le r(\t),&r\ge 0\\
                    -\tau^{z},&r(\t)\le\ve_0(\bk)\le 0,&r\le 0
                    \end{array}
             \right.
\eeqr

Working through the energy functional of Eq.
(\ref{spinful-E-functional}) in the same way as before one obtains
for $m\ne0$

\beq {\cE\over N_0}={r_0^2\over 4}(1+Z_m) \label{pom-spinful}\eeq

where the extra factor of two inside the brackets compared to Eq.
(\ref{pom-spinless}) comes from the extra trace over the
two-dimensional spin space, combined with the fact that in our
normalization, $N_0$ is the {\it spinless} single-particle density
of states. Thus, in the spinful case, with our normalizations, the
instability occurs for $\Phi_m,\ Z_m\le-1$.  From the values
quoted\cite{kwon-ceperley} in the last section, we see that the
electron gas in the clean bulk limit is quite far from these
instabilities even for $r_s=5$.

\subsection{The Disordered Bulk}
\label{disordered-bulk}

Let us now examine the changes that occur when elastic impurity
scattering is taken into account in the bulk. The RG method can be
conceptually generalized for this case as well. An important
energy scale here is $E_{\tau}=\hbar/\tau=\hbar v_F/l$, where $l$
is the mean free path due to impurities and $\tau$ the mean free
time. In a time-scale $\tau$ momentum states get scattered by
impurities implying that $E_\tau$ is the spectral broadening of a
momentum state. Since impurity scattering is elastic, an exact
eigenstate of the disordered single-particle Hamiltonian with
energy $\ve$ can be expressed as a superposition of momentum
states of energy $\ve$ with a spread of $E_{\tau}$. Therefore for
high energies $\ve\gg E_{\tau}$ integrating out a momentum states
of  thickness  $E_{\tau}$ is roughly the same as integrating out a
corresponding shell of exact energy eigenstates. This is
equivalent to the statement that states of high energy have
negligible correlations with low-energy states.


One can now ask how the interaction coupling constants flow in the
disordered case. Consider first the case of a microscopic
short-ranged interaction of range $\xi$, such as in $He^3$. The
time taken by a single fermion-fermion collision is then roughly
$\xi/v_F$. As long as this time is much smaller than $\tau$, the
collisions happen far from an impurity, and therefore conserve
total momentum. Thus, despite the fact that momentum is not a good
quantum number for single-particle states, interactions continue
to conserve momentum.  Since integrating out momentum states is
roughly the same as integrating out exact energy eigenstates, the
conclusion is that up to an energy cutoff $\Lambda$ of the order
of magnitude of $E_{\tau}$ the RG flow should proceed much as in
the clean system, resulting in Fermi liquid interactions.

Once the cutoff reaches $E_{\tau}$ integrating out momentum states
is no longer approximately equivalent to integrating out exact
eigenstates of the disordered single-particle Hamiltonian, since
now the wavefunctions of the states being integrated out have
non-negligible correlations (induced by disorder) with states
being kept. In the disorder-averaged version of the
theory\cite{int+disorder,belitz}, these correlations manifest
themselves as collective modes such as diffusons and Cooperons
with singular propagators. In the diffusive limit, these
propagators drive the RG flow\cite{int+disorder,belitz}, resulting
in a runaway flow of the $s$-wave triplet coupling $Z_0$ towards
strong coupling. The resulting state seems to be weakly
ferromagnetic\cite{weak-ferro}, but is currently not fully
understood.

Now consider a model with Coulomb interactions. Up to an energy
scale of $E_{\tau}$ a good model for the effective interactions is
the Thomas-Fermi static screened interaction $v_{TF}(q)=2\pi
e^2/(q+q_{TF})$, with a screening wavevector $q_{TF}\approx
a_0^{-1}$ in two dimensions ($a_0$ is the effective Bohr radius in
the material). By following the previous logic, it is clear that
as long as $1/q_{TF}l$ is small, electron-electron collisions will
conserve total momentum, leading to Fermi liquid interactions near
$E_F$. However, at very low energies, deep in the diffusive regime
($\omega \tau\ll1$, $ql\ll 1$) this interaction can get
unscreened\cite{altshuler-aronov}. The resulting low-energy
unscreened interaction leads to a breakdown of Fermi liquid theory
in two dimensions at the most fundamental level, namely, the
quasiparticle lifetime broadening becomes comparable to its
energy\cite{altshuler-aronov}. It is also clear that in this
regime, the electron-electron collision time is long compared to
$\tau$, and therefore momentum will not be conserved in an
electron-electron collision.

\subsection{Ballistic Quantum Dots}
\label{flt-ballistic}

Now let us examine the case of ballistic QD's, which is
conceptually the same as considering electrons in a cavity with
hard walls. Here since the role of impurity scattering is being
played by scattering from the walls of the cavity, $L$ plays the
role of $l$. Therefore the Thouless scale $E_T$ is roughly the
same as $E_{\tau}$. Since we know that the interaction is screened
at this energy scale, we can conclude that Fermi liquid
interactions are valid at this scale. Can Fermi liquid
interactions become invalid at lower energies due to
Altshuler-Aronov-like effects\cite{altshuler-aronov}? The answer
is no: Due to the finite size of the cavity, all momenta can only
be defined up to an uncertainly of $2\pi/L$. Since $q$ can only be
defined up to order $1/L$, one can never achieve the condition
$qL\ll1$. The deep diffusive regime where interactions get
unscreened is inaccessible in ballistic QD's, and is superceded by
the Random Matrix (RMT) regime which we will analyze in detail in
the next section.  The conclusion of this set of arguments is that
Fermi liquid interactions are natural for ballistic QD's, while
they may not be for strongly diffusive ($l\ll L$) QD's.

Let us now explicitly consider what the low-energy effective
Hamiltonian looks like for a ballistic QD in the Thouless band. We
will first write it down, and then discuss it.

\beq H=\sum\limits_{\a}\ve_\a c^{\dagger}_\a c_\a
+{\Delta\over2}\sum\limits_{\bk,\bk'} u(\t-\t'):n_\bk n_{\bk'}:
\label{hball-nospin}\eeq

where $\a$ are labels for the exact eigenstates ($g$ in number) of
the single-quasiparticle Hamiltonian encoding the chaotic boundary
scattering, $\Delta$ (the mean level spacing) is the inverse
density of spinless states $1/N_0$,
$n_{\bk}=c^{\dagger}(\bk)c(\bk)$, and it is understood that the
sum over $\bk,\ \bk'$ goes over $g$ values near the Fermi surface.
(We are using  $n_\bk$ instead of  $\delta n_\bk$, the difference
being absorbed in  a shift in the chemical potential.)  The
Hamiltonian is expressed in a hybrid basis, with the connection
between the $\a$ and the $\bk$ basis being given by the
wavefunctions of the exact eigenstates $\a$ in the momentum basis
$\phi_\a(\bk)$. The statistics of the energies $\ve_\a$ and
wavefunctions $\phi_\a(\bk)$ are assumed to be controlled by RMT.
Note that the momentum $\bk$ is uncertain up to $2\pi/L$, and it
really represents a patch on the Fermi surface. To be more
precise, we can define the state we label by $\bk$ in the dot as
\beq |\bk\rangle=\sum\limits_{\a} |\a\rangle\langle
\a|e^{i\bk\cdot\br}\rangle=\sum\limits_{\a} |\a\rangle
\phi_{\a}(\bk) \label{defn-of-k} \eeq For convenience we will
choose the $\bk$ labels to be equally spaced (separated by an angle
$2\pi/k_FL=2\pi/g$) on the Fermi circle. This set of states clearly
satisfies all the boundary conditions of the dot, and is unitarily
related to the set of disorder eigenstates $|\a\rangle$. In the
original Fermi liquid Hamiltonian (Eq.  (\ref{hfl})) there was an
additional sum over $\bq$. Here, since $\bq$ is also uncertain up to
$2\pi/L$ and only scattering within the Thouless shell is allowed, the
sum over $\bq$ is eliminated.  Since the $\bk$ take on $g$ discrete
values, Landau parameters with angular momentum $m$ of order $g$ are
not sensible in this theory. Our focus will be on small $m$ of order
1, because these are the channels in which the Landau parameters are
expected to have large magnitudes.

An interesting fact about ballistic/chaotic quantum dots is that the
ratio $E_F/E_T$ is related to $g$:
\beq
{E_F\over E_T}={\hbar^2 k_F^2/2m^*\over \hbar v_F/L}={k_FL\over2}\simeq g
\eeq
Recall that Fermi liquid theory becomes applicable only in an energy
shell of size much smaller than $E_F$ around the Fermi
surface. Therefore the largeness of $g$ is also a sufficient condition
for the validity of the Fermi liquid interactions.

To summarize, in writing down the Hamiltonian of
Eq. (\ref{hball-nospin}), we are appealing to the fact that whereas
single particle momentum is not conserved due to collision with the
walls, the momentum of a pair is conserved during a collision if it
takes place over a time scale shorter than $L/v_F$, the time to bounce
off the wall. The interacting part of the Hamiltonian continues to
have the same form as in the clean limit even in the presence of
chaotic boundary scattering.

\subsection{Recovering the Universal Hamiltonian}
\label{flt-univ-ham}

Let us see  how  the Universal hamiltonian of Eqn. (\ref{hu})
emerges, starting with the spinless case.  One can use the
$\phi_\a(\bk)$ to express the entire Hamiltonian in the $\a$ basis
in the form of Eq. (\ref{hran}), with

\begin{eqnarray}
V_{\alpha \beta \gamma \delta}=& {\Delta\over 4}\sum\limits_{\bk
\bk'}u (\theta -\theta' ) \left[ \phi^{*}_{\alpha}(\bk )
\phi^{*}_{\beta}(\bk' ) -\phi^{*}_{\alpha}(\bk' )
\phi^{*}_{\beta}(\bk )\right] \nonumber\\ &\times \left[
\phi_{\gamma}(\bk' ) \phi_{\delta}(\bk ) -\phi_{\gamma}(\bk )
\phi_{\delta}(\bk' )\right] \label{V-disorder-basis}\end{eqnarray}

These coefficients $V_{\alpha \beta \gamma \delta}$ will vary from
sample to sample. Let us focus on terms that survive the ensemble
average, which here reduces to a Random Matrix Theory\cite{RMT}
average. Let us concentrate on the GOE, where all the
wavefunctions in the real-space basis $\phi_\a(\br)$ can be made
real. Since we are working in an approximate momentum basis
labeled by the $g$ values of $\bk$, the reality of the real-space
wavefunctions translates to

\beq \big(\phi_\a(\bk)\big)^*=\phi_\a(-\bk)
\label{goe-k-minus-k-connection}\eeq

The fundamental RMT wavefunction correlator is\cite{RMT}

\beq
<\phi_\a^*(\bk)\phi_\b(\bk')>={\delta_{\a\b}\delta_{\bk\bk'}\over
g} \eeq

To find the average of $V_{\alpha \beta \gamma \delta}$ we need
the four-point correlator.  Apart from exceptional cases when all
the subscripts and/or all the arguments of the wavefunctions are
equal, the four-point wavefunction correlator in the GOE can be
written to leading order in $1/g$ as

\beqr
<\phi_\a^*(\bk_1)\phi_\b^*(\bk_2)\phi_\g(\bk_3)\phi_\d(\bk_4)>&=
{\d_{\a\d}\d_{\b\g}\d_{\bk_1\bk_4}\d_{\bk_2\bk_3}\over
g^2}\nonumber\\
&+{\d_{\a\g}\d_{\b\d}\d_{\bk_1\bk_3}\d_{\bk_2\bk_4}\over
g^2}\nonumber\\
&+{\d_{\a\b}\d_{\g\d}\d_{\bk_1,-\bk_2}\d_{\bk_3,-\bk_4}\over g^2}
\label{4pt1} \eeqr

It is seen that only matrix elements in which the indices $\a\b\g\d$
are pairwise equal survive disorder-averaging, and also that the
average has no dependence on the energy of $\a\b\g\d$. In the spinless
case, the first two terms on the right hand side make equal
contributions and produce the constant charging energy in the
Universal Hamiltonian of Eq. (\ref{hu}), while in the spinful case (to
be discussed at length in Section \ref{rg-spinful}) they produce the
charging and exchange terms. The final term of Eq.  (\ref{4pt1})
produces the Cooper interaction of Eq. (\ref{hu}).

Finally, one can explicitly calculate the variances

\beq \langle V^{2}_{\alpha \beta \gamma \delta }\rangle - \langle
V_{\alpha \beta \gamma \delta }\rangle^{2}={\Delta^2 \over
4g^2}\sum_{m=1}u_{m}^{2}.\label{disp} \eeq and see that they are
small. Note that $u_0$ does not contribute to the fluctuations
between different disorder realizations.

\section{Instabilities of the Universal Hamiltonian: Renormalization Group Treatment}
\label{rg-mm}

The Universal Hamiltonian\cite{H_U,univ-ham} is obtained by replacing
the interaction by its ensemble average in the full Hamiltonian,
arguing that sample-to-sample fluctuations in the interaction matrix
elements are small. At this point one can ask when small terms in the
Hamiltonian can be safely discarded, and when they
cannot\cite{qd-us1}. As long as one is interested in low-energy
properties, the RG is the perfect tool to answer this question. One
simply integrates out high energy states, and looks at the fate of the
originally small couplings. If they are irrelevant in the RG sense,
discarding them is justified, while if they are relevant they dominate
the low-energy physics regardless of how small they were initially.

Two of us carried out just such an RG analysis\cite{qd-us1}, to the
presentation of which this section is devoted. We will first consider
the spinless case, and then make some remarks about the spinful case. We
will close this section with a list of questions left unanswered by
the RG.

\subsection{RG for the Spinless Case}
\label{rg-spinless}

We start with the Hamiltonian for spinless fermions in a ballistic
QD, Eq. (\ref{hball-nospin}), which we reproduce here for
convenience

\beq H=\sum\limits_{\a}\ve_\a c^{\dagger}_\a c_\a
+{\Delta\over2}\sum\limits_{\bk,\bk'} u(\t-\t'):n_\bk n_{\bk'}:
\label{hball-nospin2} \eeq

\begin{figure}
\narrowtext
\epsfxsize=2.4in\epsfysize=2.0in
\hskip 0.3in\epsfbox{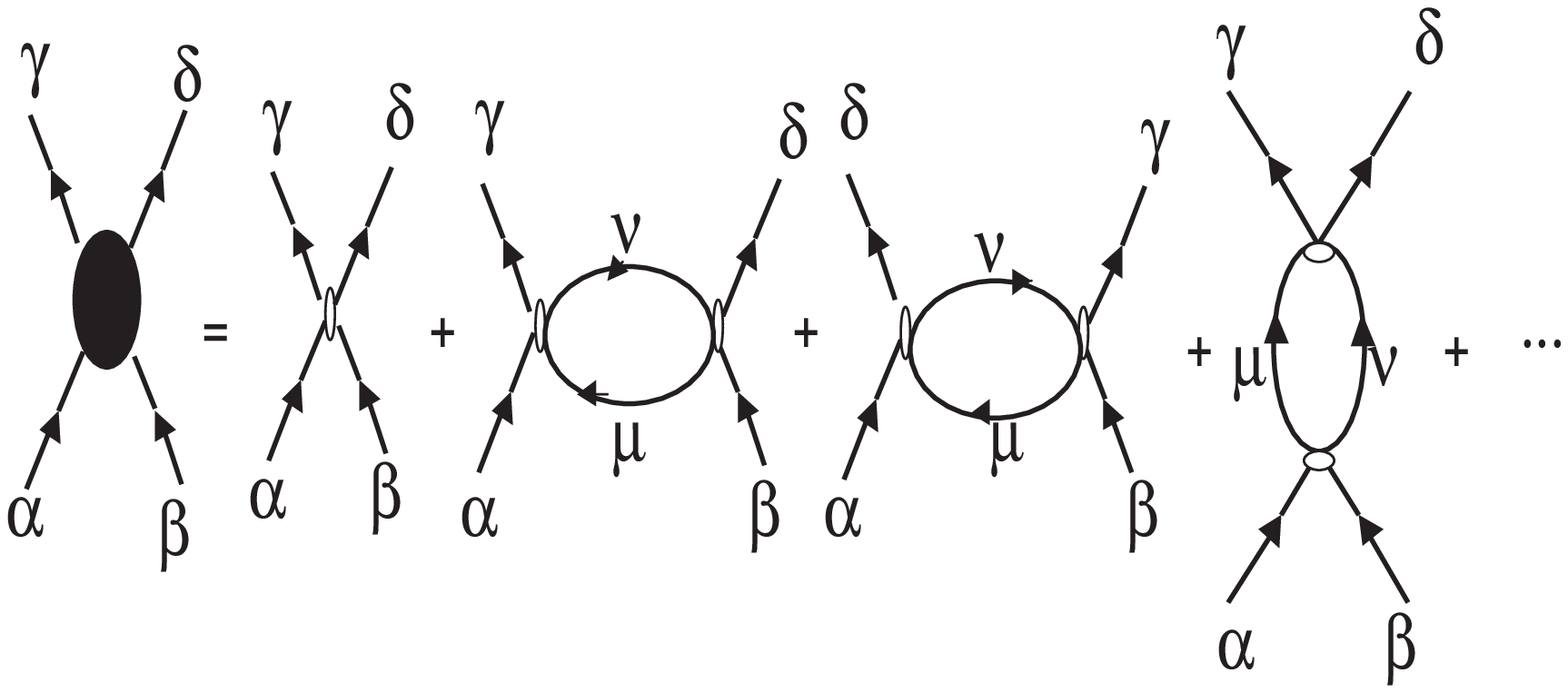}
\vskip 0.15in
\caption{Feynman diagrams for the full four-point amplitude $\Gamma_{\a\b\g\d}$.}
\label{scatt-amplitude1}
\end{figure}

The strategy for carrying out RG in a finite system\cite{rg-us} is the
following: (i) Since we are in the Thouless shell we cannot integrate
out momenta, but must integrate out exact eigenstates of the chaotic
single-particle Hamiltonian.  (ii) After integrating out some exact
eigenstates at a given stage in the RG we have $g'=ge^{-\xi}$ levels
left (here $\xi$ is called the flow parameter of the RG). At this
stage we compute a scattering amplitude $\Gamma_{\a\b\g\d}$ for the
process in which two fermions originally in states $\a\b$ are
scattered into states $\g\d$. This scattering can proceed directly
through the vertex $V_{\a\b\g\d}(\xi)$, or via intermediate virtual
states higher order in the interactions, which can be classified by a
set of Feynman diagrams, as shown in Figure 1. All the states in the
diagrams belong to the $g'$ states kept.  (iii) We demand that the
entire amplitude be independent of $g'$, meaning that the physical
amplitudes should be the same in the effective theory as in the
original theory. This will lead to a set of flow equations for the
$V_{\a\b\g\d}$. In principle this flow equation will involve all
powers of $V$ but we will keep only quadratic terms (the one-loop
approximation). Then the diagrams are limited to the ones shown in
Figure 1, leading to the following contributions to the scattering
amplitude $\Gamma_{\a\b\g\d}$

\beqr \Gamma_{\a\b\g\d}=&V_{\a\b\g\d}\nonumber \\
+&\sum\limits_{\mu,\nu}^{} {}'
{N_F(\nu)-N_F(\mu)\over\ve_{\mu}-\ve_{\nu}}
(V_{\a\nu\mu\d}V_{\b\mu\nu\g}-V_{\a\nu\mu\g}V_{\b\mu\nu\d})\nonumber
\\ -&\sum\limits_{\mu\nu}^{}
{}'{1-N_F(\mu)-N_F(\nu)\over\ve_{\mu}+\ve_{\nu}}
V_{\a\b\mu\nu}V_{\nu\mu\g\d} \label{rgflow}\eeqr where the prime
on the sum reminds us that only the $g'$ remaining states are to
be kept and $N_F(\a)$ is the Fermi occupation of the state $\a$.
We will confine ourselves to zero temperature where this number
can only be zero or one, but the extension to finite temperature
is straightforward. Also recall that the matrix element
$V_{\a\b\g\d}$ now explicitly depends on the RG flow parameter
$\xi$.

Now we demand that upon integrating the two states at $\pm
g'\Delta/2$ we recover the same $\Gamma_{\a\b\g\d}$. Clearly,
since $g'=ge^{-\xi}$,

\beq {d\over d\xi}=-g'{\d\over\d g'} \label{xisign}\eeq

The effect of this differentiation on the loop diagrams is to fix
one of the internal lines of the loop to be at the cutoff $\pm
g'\Delta/2$, while the other one ranges over all smaller values of
energy. In the particle-hole diagram, since $\mu$ or $\nu$ can be
at $+g'\Delta/2$ or $-g'\Delta/2$, and the resulting summations
are the same in all four cases, we take a single contribution and
multiply by a factor of 4. The same reasoning applies to the
Cooper diagram. Let us define the energy cutoff
$\Lambda=g'\Delta/2$ to make the notation simpler. Since we are
integrating out two states we have $\d g'=2$

\beqr 0=&{dV_{\a\b\g\d}\over d\xi}\nonumber \\ -&{g'\over
2}4\sum\limits_{\mu=\Lambda,\nu}^{} {}'
{N_F(\nu)-N_F(\mu)\over\ve_{\mu}-\ve_{\nu}}
(V_{\a\nu\mu\d}V_{\b\mu\nu\g}-V_{\a\nu\mu\g}V_{\b\mu\nu\d})\nonumber
\\ +&{g'\over 2}4\sum\limits_{\mu=\Lambda,\nu}^{}
{}'{1-N_F(\mu)-N_F(\nu)\over\ve_{\mu}+\ve_{\nu}}
V_{\a\b\mu\nu}V_{\nu\mu\g\d} \label{rgflow2}\eeqr The changed sign
in front of the 1-loop diagrams reflects the sign of Eq.
(\ref{xisign})

So far we have not made any assumptions about the form of
$V_{\a\b\g\d}$, and the formulation applies to any finite system.
In a generic system such as an atom\cite{rg-us}, the matrix
elements depend very strongly on the state being integrated over,
and the flow must be followed numerically for each different set
$\a\b\g\d$ kept in the low-energy subspace. The crucial
simplification in QD's comes from the fact that these matrix
elements are random and controlled by RMT, and that the {\em  RG
equation is self-averaging}, as will be seen below.

Let us go back to the properly antisymmetrized matrix element
defined in terms of the Fermi liquid interation function, Eq.
(\ref{V-disorder-basis}). Since there is a product of two $V$'s in
each loop diagram, and each $V$ contains 4 terms, it is clear that
each loop contribution has 16 terms. However, it turns out that
only certain terms contribute to leading order in the large-$g$
limit, while others do not. We will focus on one of each type to
illustrate the difference. Let us first consider a term of type I
in the particle-hole diagram, which survives in the large-$g$
limit. Putting in the full wavefunction dependences (and ignoring
factors other than $g,\ g'$) we have

\beqr &g'\Delta^2\sum\limits_{\nu=-\Lambda}^{0}
{1\over\Lambda+|\ve_\nu|} \nonumber\\
&\sum\limits_{\bk\bk'}\sum\limits_{\bp\bp'}
u(\t_\bk-\t_\bp)u(\t_{\bp'}-\t_{\bk'})\nonumber\\
&\phi^*_\a(\bk)\phi^*_\b(\bk')\phi_\g(\bk')\phi_\d(\bk)\nonumber\\
&\phi^*_\mu(\bp)\phi^*_\nu(\bp')\phi_\nu(\bp)\phi_\mu(\bp')
\label{diagram-leading}\eeqr

The internal sum over $ \nu$ is self-averaging. While the most
convincing way to show this is to compute its variance, and see that
it is of order $1/\sqrt{g}$ times its average, this fact can be
motivated in the following way: There is a sum over $g'\gg1$ values of
$\nu$ with a slowly varying energy denominator, which makes the sum
over $\nu$ similar to a spectral average, which in RMT is the same as
an average over the disorder ensemble.  A more sophisticated argument
is presented in Appendix A.


In RMT the wavefunction averages do not depend on the energy
separations of the states, so the wavefunction average can be
carried out separately. This average in the GOE for $\mu\ne\nu$
and generic momentum labels is

\beqr
<\phi^*_\mu(\bp_1)\phi^*_\nu(\bp_2)\phi_\nu(\bp_3)\phi_\mu(\bp_4)>=\nonumber\\
{\d_{14}\d_{23}\over g^2}-{\d_{13}\d_{24}\over
g^3}-{\d_{1,-2}\d_{3,-4}\over g^3} \label{crucial1}\eeqr

The $1/g^2$ term is the ``naive Wick contraction'' of
leading-order RMT, but the $1/g^3$ terms are necessary to maintain
orthogonality between $\mu$ and $\nu$. The final term would be
missing in the GUE. Substituting the correct momentum labels for
the particle-hole diagram we see that the wavefunction average is

\beq {\d_{\bp\bp'}\over g^2}-{1+\d_{\bp,-\bp'}\over g^3}
\label{crucial2}\eeq

Using the self-averaging shown in Appendix A, the first term of
Eq.(\ref{crucial2}) forces $\bp=\bp'$ in Eq.
(\ref{diagram-leading}). For large $g$, using \beq
\sum\limits_{\bp} =g\int {d\t_\bp\over 2\pi} \eeq we obtain  a
convolution of the two Fermi liquid functions \beq
\sum\limits_{\bp}u(\t_\bk-\t_\bp)u(\t_\bp-\t_{\bk'})=
g\big(u_0^2+\half\sum\limits_{m=1}^{\infty}
u_m^2\cos{m(\t-\t')}\big) \label{convolve}\eeq where we have
reverted to the notation $\t=\t_\bk,\ \t'=\t_{\bk'}$. In the
second term of Eq. (\ref{crucial2}), the $\d_{\bp,-\bp'}$ turns
out to be subleading, while the other allows independent sums over
$\bp,\ \bp'$. This means that only $u_0$ contributes due to this
term, which produces

\beq
\sum\limits_{\bp\bp'}u(\t_\bk-\t_\bp)u(\t_{\bp'}-\t_{\bk'})=g^2
u_0^2 \label{convolve-u0}\eeq

We still need to do the energy sum, which we replace by an
integral \beq
\sum\limits_{\ve_\nu=-\Lambda}^{0}{1\over\Lambda+|\ve_\nu|}\approx
\int\limits_{0}^{\Lambda}{d\ve\over\Delta}{1\over
\Lambda+\ve}={\ln{2}\over\Delta} \eeq Feeding this into full
expression for this contribution to the particle-hole diagram, we
find it to be \beqr &{g'\over g}\Delta
\ln{2}\sum\limits_{\bk\bk'}\bigg(\sum\limits_{m=1}^{\infty}
u_m^2\cos{m(\t-\t')}\bigg)\nonumber\\
&\phi^*_\a(\bk)\phi^*_\b(\bk')\phi_\g(\bk')\phi_\d(\bk) \eeqr

Notice that the result is still of the Fermi liquid form.  It is
also seen that the contribution from $u_0$ cancels, about which we
will say more below.

Now let us consider a contribution which is subleading in $1/g$ in
the large-$g$ limit.

\beqr &-g'\Delta^2\sum\limits_{\nu=-\Lambda}^{0}
{1\over\Lambda+|\ve_\nu|} \nonumber\\
&\sum\limits_{\bk\bk'}\sum\limits_{\bp\bp'}
u(\t_\bk-\t_\bp)u(\t_{\bp'}-\t_{\bk'})\nonumber\\
&\phi^*_\a(\bp)\phi^*_\b(\bk')\phi_\g(\bk')\phi_\d(\bk)\nonumber\\
&\phi^*_\mu(\bk)\phi^*_\nu(\bp')\phi_\nu(\bp)\phi_\mu(\bp')
\label{diagram-subleading}\eeqr

Note that the momentum labels of $\phi^*_\a$ and $\phi^*_\mu$ have
been exchanged compared to Eq. (\ref{diagram-leading}) and there
is a minus sign, both coming from the antisymmetrization of Eq.
(\ref{V-disorder-basis}). Once again we ensemble average the
internal $\mu,\ \nu$ sum, the wavefunction part of which gives

\beqr
<\phi^*_\mu(\bk)\phi^*_\nu(\bp')\phi_\nu(\bp)\phi_\mu(\bp')>=\nonumber\\
{\d_{\bk\bp'}\d_{\bp\bp'}\over
g^2}-{\d_{\bk\bp}+\d_{\bk,-\bp'}\d_{\bp,-\bp'}\over g^3} \eeqr

It is clear that there is an extra momentum restriction in each
term compared to Eq. (\ref{crucial2}), which means that one can no
longer sum freely over $\bp$ to get the factor of $g$ in Eq.
(\ref{convolve}), or the factor of $g^2$ in Eq.
(\ref{convolve-u0}). Therefore this contribution will be down by
$1/g$ compared to that of Eq. (\ref{diagram-leading}). The general
rule is that whenever a momentum label corresponding to an
internal line in the diagram (here $\mu$ and $\nu$) is forced to
become equal to a momentum label corresponding to an external
disorder label (here $\a , \b , \g , $ or $\d$), the diagram is
down by $1/g$, exactly as in the $1/N$ expansion.

Turning now to the Cooper diagrams, the internal lines are forced
to have the same momentum labels as the external lines by the
Fermi liquid vertex, therefore they do not make any leading
contributions.

Finally, one can collect all the terms, and conclude that for
$m\ne0$ \beq {du_m\over d\xi}=-e^{-\xi}(\ln{2})u_m^2
\label{rg-spinless1}\eeq

It must be emphasized that $u_0$ does not flow. The physical
reason for this is that it commutes with the one-body ``kinetic''
part, and therefore does not suffer quantum fluctuations. The
above equation can be written in a more physically transparent
form by using a rescaled (for $m\ne0$ only) variable

\beq \tu_m=e^{-\xi} u_m \eeq

in terms of which the flow equation becomes

\beq {d\tu_m\over d\xi}=-\tu_m-(\ln{2})\tu_m^2\equiv \beta(\tu_m)
\label{rg-spinless2}\eeq where the last is a definition of the
$\b$-function.  To understand the meaning of $\tu_m$ we go back to
Eq. (\ref{disp}), which expresses the variances of $V_{\a\b\g\d}$
in terms of $u_m$'s. We start with an ensemble of QD's with
dimensionless conductance $g$. Suppose we wish to define a new
ensemble with dimensionless conductance $g'$ and Fermi liquid
parameters $u_m'$, and demand that {\it all the statistical
properties of the matrix elements, including the variances of Eq.
(\ref{disp}), be the same}, we are led to

\beqr u_0'=u_0\\ {u_m'^2\over g'^2}={u_m^2\over g^2}\Rightarrow
u_m'=e^{-\xi} u_m \eeqr

It is clear that the $u_m'$ are exactly the $\tu_m$. From Eq.
(\ref{rg-spinless2}) it can be seen that positive initial values
of $\tu_m$ (which are equal to initial values of $u_m$ inherited
from the bulk) are driven to the fixed point at $\tu_m=0$, as are
negative initial values as long as $u_m(\xi=0)\ge
u_m^*=-1/\ln{2}$. Thus, the Fermi liquid parameters are {\it
irrelevant} for this range of starting values. Recall that setting
all $u_m=0$ for $m\ne0$ results in the Universal
Hamiltonian\cite{H_U}. Thus, the range $u_m\ge u_m^*$ is the basin
of attraction of the Universal Hamiltonian.  On the other hand,
for $u_m(\xi=0)\le u_m^*$ results in a runaway flow towards large
negative values of $u_m$, signalling a phase transition to a phase
not perturbatively connected to the Universal Hamiltonian. Recall
that the Pomeranchuk instability\cite{pomeranchuk} of the clean
spinless Fermi liquid happens for $u_m\le -2$. The flow towards
large negative values of $u_m$ suggests that the system is
undergoing some kind of Pomeranchuk instability. This notion turns
out to be correct, and will be made precise in the next section.
Note that there is a window $-2\le u_m\le u_m^*$ in which the
clean bulk system is stable while the system in a ballistic QD is
unstable.

Let us summarize the results of the RG: (i) The RG equation is
self-averaging for large $g$. (ii) The Fermi liquid form of the
interactions is left unchanged to leading order by the RG flow.
(iii) Each $u_m$ flows separately, while $u_0$ does not flow. (iv)
The flow equations show an instability for $u_m(\xi=0)\le u_m^*=
-1/\ln{2}$ in every Fermi liquid channel, presumably towards a
Pomeranchuk-like state. (v) The critical point shows a
``correlation length'' exponent $\nu=1$, calculated from the slope
of the $\b$-function at the critical point.

\beq \nu=\bigg({d\b\over d\tu_m}\bigg)_{u_m^*} =1\eeq

\subsection{Some Remarks on the Spinful Case}
\label{rg-spinful}

The case with spin is similar conceptually but somewhat more
complicated technically than the spinless case. For a
spin-rotationally-invariant Hamiltonian, the two-body interaction
can be decomposed\cite{landau,agd} into a spin-singlet interaction
function $u^{(s)}(\t-\t')$ and a spin-triplet interaction function
$u^{(t)}(\t-\t')$. The projectors for the two two-body spin states
are

\beqr \cP^{(s)}={1-\vtau\cdot\vtau'\over4}\\
\cP^{(t)}={3+\vtau\cdot\vtau'\over4} \eeqr where $\vtau$ and
$\vtau'$ represent the Pauli spin operators for the two electrons
(the numbers 1 and 3 in the numerators should be thought of as
unit matrices in spin space).  Writing the total interaction and
comparing it to the Fermi liquid form (Eq. (\ref{fli-spin2})) we find

\beq u^{(s)}\cP^{(s)}+u^{(t)}\cP^{(t)}=\Phi + Z \vtau\cdot\vtau'
\eeq

where all Fermi liquid labels and spin-matrix indices have been
suppressed. This leads to the following relations \beqr
\Phi={u^{(s)}+3u^{(t)}\over 4}\\ Z={u^{(t)}-u^{(s)}\over 4} \eeqr

In order to carry out the RG, we define the interaction
Hamiltonian as \beq H_I=\half\sum\limits_{ss',\a\b\g\d}
V^{ss'}_{\a\b\g\d} c^{\dagger}_{\a s}c^{\dagger}_{\b s'}c_{\g
s'}c_{\d s} \eeq

The one-loop scattering amplitude now reads as follows

\beqr &\Gamma^{ss'}_{\a\b\g\d}=V^{ss'}_{\a\b\g\d}\nonumber\\
&+\sum\limits_{\mu\nu}{N_F(\nu)-N_F(\mu)\over\ve_\mu-\ve_\nu}\bigg(V^{ss'}_{\a\mu\g\nu}V^{s's}_{\b\nu\d\mu}-V^{ss}_{\a\mu\nu\d}V^{s's}_{\b\nu\mu\g}\nonumber\\&-V^{s's'}_{\b\nu\mu\g}V^{ss'}_{\a\mu\nu\d}-\sum_{s_1}V^{ss_1}_{\a\mu\nu\d}V^{s's_1}_{\b\nu\mu\g}\bigg)\nonumber\\
&-\sum\limits_{\mu\nu}{1-N_F(\mu)-N_F(\nu)\over\ve_\mu+\ve_\nu}V^{ss'}_{\a\b\mu\nu}V^{ss'}_{\nu\mu\g\d}
\eeqr

In order to separate $V^{ss'}_{\a\b\g\d}$ into its singlet and triplet
components, we need two facts: (i) The singlet channel is
characterized by antisymmetry in spin, and therefore symmetry in the
orbital labels of the incoming (and separately, the outgoing)
particles. The triplet channel is exactly the reverse. (ii) If $s=s'$
then the interaction is forced to be in the triplet channel. However,
if $s\ne s'$ the interaction is a superposition of the singlet and
triplet channels.

On the above basis we identify the singlet and triplet channel
interactions as

\beqr
V^{(s)}_{\a\b\g\d}=&{1\over4}\bigg(V^{\ua\da}_{\a\b\g\d}+V^{\ua\da}_{\b\a\g\d}+V^{\ua\da}_{\a\b\d\g}+V^{\ua\da}_{\b\a\d\g}\bigg)\\
V^{(t)}_{\a\b\g\d}=&{1\over4}\bigg(V^{\ua\da}_{\a\b\g\d}-V^{\ua\da}_{\b\a\g\d}-V^{\ua\da}_{\a\b\d\g}+V^{\ua\da}_{\b\a\d\g}\bigg)=V^{\ua\ua}_{\a\b\g\d}
\eeqr

and the inverse relations

\beqr V^{\ua\ua}_{\a\b\g\d}=V^{(t)}_{\a\b\g\d}\\
V^{\ua\da}_{\a\b\g\d}=V^{(s)}_{\a\b\g\d}+V^{(t)}_{\a\b\g\d} \eeqr

In order to carry out the wavefunction averages one expresses
these matrix elements as

\beqr V^{(s)}_{\alpha \beta \gamma \delta}=& {\Delta\over
4}\sum\limits_{\bk \bk'}u^{(s)} (\theta -\theta' ) \left[
\phi^{*}_{\alpha}(\bk ) \phi^{*}_{\beta}(\bk' )
+\phi^{*}_{\alpha}(\bk' ) \phi^{*}_{\beta}(\bk )\right]
\nonumber\\ &\times \left[ \phi_{\gamma}(\bk' ) \phi_{\delta}(\bk
) +\phi_{\gamma}(\bk ) \phi_{\delta}(\bk' )\right] \\
V^{(t)}_{\alpha \beta \gamma \delta}=& {\Delta\over
4}\sum\limits_{\bk \bk'}u^{(t)} (\theta -\theta' ) \left[
\phi^{*}_{\alpha}(\bk ) \phi^{*}_{\beta}(\bk' )
-\phi^{*}_{\alpha}(\bk' ) \phi^{*}_{\beta}(\bk )\right]
\nonumber\\ &\times \left[ \phi_{\gamma}(\bk' ) \phi_{\delta}(\bk
) -\phi_{\gamma}(\bk ) \phi_{\delta}(\bk' )\right] \eeqr

Now it is straightforward to find the RG flows of the singlet and
triplet Landau interaction functions. A nontrivial check is
provided by the fact that there are two expressions for the
triplet interaction, and their flows should be identical. These
flows can be written in terms of the rescaled Landau parameters
${\tilde\Phi}_m=e^{-\xi}\Phi_m$ and ${\tilde Z}_m=e^{-\xi}Z_m$,
and one obtains to leading order in $1/g$

\beqr {d{\tilde \Phi}_m\over
d\xi}&=-{\tilde\Phi}_m-(2\ln{2}){\tilde\Phi}_m^2\\ {d{\tilde
Z}_m\over d\xi}&=-{\tilde Z}_m-(2\ln{2}){\tilde Z}_m^2 \eeqr

Clearly, the charge and spin channels flow independently, and
there is an instability at $-1/(2\ln{2})=-0.7213\dots$ in each
channel. Recall that in the clean Fermi liquid the Pomeranchuk
instability occured at $-1$ in our normalization. Thus, there is a
substantial window where the clean system in the bulk is stable,
but the ballistic system is unstable.

\subsection{Unanswered Questions in the RG}
\label{rg-not-answered}

While the RG makes it plausible that there is indeed a transition
out of the phase controlled by the Universal Hamiltonian, it
leaves two important questions unanswered. Recall that we have
used a one-loop calculation of the scattering amplitude to define
the RG flow. This is usually sufficient when the dimensionless
couplings $u_m$ are much smaller than unity. However, the critical
point occurs when $u_m$ is of order unity. Can we trust the
one-loop calculation at this point? In other words, suppose there
were higher-order diagrams in the scattering amplitude to leading
order in $1/g$. This would have generated cubic and higher-order
terms in the RG flow. For a sufficiently large value of the cubic
term, the critical point identified in the previous subsections
can disappear. In that case it would have been an artifact based
on an invalid approximation. Secondly, even if we accept the phase
transition as genuine, the RG gives us no way of calculating
physical quantities in the strong-coupling phase. Both of these
questions are answered in the next section, where we show that the
one-loop flow is exact in the large-$g$ limit, and that there is a
controlled way of calculating all the properties of the
strong-coupling phase in a large-$N$ approximation.

\section{Large-$N$ Theory of the Mesoscopic Pomeranchuk Regimes}
\label{large-N}

As the name suggests, large-$N$ theories typically involve
interactions between $N$ species of objects. The largeness of $N$
renders fluctuations (thermal or quantum) small, and enables one
to make approximations which are not perturbative in the coupling
constant, but are controlled by the additional small parameter
$1/N$. Large-$N$ approximations have found many uses in condensed
matter physics, but usually the value of $N$ which characterizes
the real model is quite small (2 or 3). We will find below that
ballistic mesoscopic structures offer a realization of large-$N$
theory where $N=g$. The $g$ momentum states act as $g$ ``species''
of fermions. Thus, one can make $N$ as large as one wants modulo
technological challenges. In fact, standard Fermi Liquid theory in
a clean bulk system can also be formulated in this
way\cite{rg-shankar}.

There are two primary manifestations of the large-$N$ nature of a
theory: (i) The four-point scattering amplitude is dominated by
iterated particle-hole diagrams. (ii) The effective action is
dominated by its saddle-point.

In this section we will demonstrate each of these manifestations
of large-$N$ theory, and end by solving the strong-coupling theory
in a saddle-point approximation.

\subsection{Diagrammatic Large-$N$ and the Exactness of One-Loop RG in the Large-$g$
Limit} \label{large-N-diagrams}

To put the issue in context, let us consider the Gross-Neveu
model\cite{gross-neveu} which is one of the simplest fermionic
large-$N$ theories. This theory has $N$ identical massless
relativistic fermions interacting by an attractive short-range
interaction. The Lagrangian density is

\beq \cL=\sum\limits_{i}{\bar \psi}_i\not\!\partial \psi_i
-{\lambda\over N} \bigg(\sum\limits_{i}{\bar\psi}_i\psi_i\bigg)^2
\eeq

Note that the propagator conserves the internal index, as does the
interaction term.

\begin{figure}
\narrowtext \epsfxsize=2.4in\epsfysize=2.0in \hskip
0.3in\epsfbox{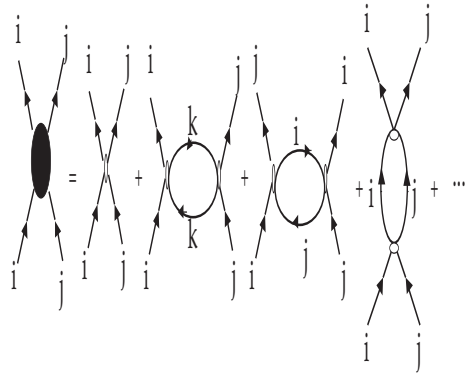} \vskip 0.15in
\caption{ The first few
Feynman diagrams in the four-point scattering amplitude for the
Gross-Neveu theory.}
\label{scatt-amplitude-2}
\end{figure}

Figure 2 shows the first few diagrams in the expression for the
four-point scattering amplitude (for particles of species $i$ and
$j$ to scatter) in the Gross-Neveu theory. The ``bare'' vertex
comes with a factor $\lambda/N$. The one-loop diagrams all share a
factor $\lambda^2/N^2$ from the two vertices. The first one-loop
diagram has a free internal summation over the index $k$ over $N$
values, with the contribution being identical for each value of
$k$. Thus, this one-loop diagram acquires a compensating factor of
$N$ which makes its contribution of order $\lambda^2/N$, the {\it
same order in $1/N$ as the bare vertex}. However, the other
one-loop diagrams have no such free internal summation and their
contribution is indeed of order $1/N^2$.Therefore, to leading
order in $1/N$, one should keep only diagrams which have a free
internal summation for every vertex, that is, iterates of the
leading one-loop diagram, which are called bubble graphs. {\em For
later use remember that in the diagrams that survive (do not
survive), the indices $i$ and $j$ of the incoming particles do not
(do) enter the loops.}  Let us assume that the momentum integral
up to the cutoff $\Lambda$ for one bubble gives a factor
$\Pi(\Lambda, q_{ext})$, where $q_{ext}$ is the external momentum
transfer at which the scattering amplitude is evaluated. To
leading order in large-$N$ the full expression for the scattering
amplitude is

\beq \Gamma_{ij}(q_{ext})={1\over N}{\lambda\over1+\lambda
\Pi(\Lambda, q_{ext})} \eeq

Once one has the full expression for the scattering amplitude (to
leading order in $1/N$) one can ask for the RG flow of the
``bare'' vertex as the cutoff is reduced by demanding that the
physical scattering amplitude $\Gamma$ remain insensitive to the
cutoff. One then finds

\beq {d\Gamma_{ij}(q_{ext})\over d\xi}=0\Rightarrow {d\lambda\over
d\xi}=\lambda^2 {d\Pi(\Lambda, q_{ext})\over d\xi} \eeq where
$\xi=\ln(\Lambda_0/\Lambda)$ is the flow parameter of the RG. This
equation shows that {\it the one-loop RG flow is the exact answer
to leading order in a large-$N$ theory}. All higher-order
corrections must therefore be subleading in $1/N$.

Let us now turn to our theory, in which the role of the internal
index is played by the momentum patch label $\bk$, which can take
on $g$ values. The role of the coupling constant $\lambda$ in our
theory is played by $E_T u(\t-\t')$. From the definition of the
matrix elements in Eq. (\ref{V-disorder-basis}) we see that
interactions occur in the Hamiltonian as $E_T u(\t-\t')/g$ in
analogy with the $\lambda/N$ in the Gross-Neveu Lagrangian. Let us
remark in this context that the most natural way to take the
large-$g$ limit is to keep $E_T$ fixed and decrease $\Delta$.

The major departure from the Gross-Neveu theory is that in our
theory the propagator does not conserve the label $\bk$. The
contributions of the leading and subleading one-loop diagrams were
presented in Eqs. (\ref{diagram-leading},\ref{diagram-subleading})
respectively. The crucial fact which allows us to show the
large-$N$ nature of this theory is the self-averaging of the
internal summations, which follows from the arguments of the
previous section (and Appendix A). This self-averaging means that
the internal wavefunction products can be replaced by their
ensemble averages, which by Eq. (\ref{crucial1}) makes the
internal propagators effectively momentum-conserving. Now the
analogy to the Gross-Neveu theory is complete.

>From this analogy we learn that the one-loop RG flow computed by two
of us\cite{qd-us1} is indeed the exact answer in the large-$g$
limit, and that the phase transition it indicates is real.
However, this analogy is even more fruitful in providing a way to
compute physical quantities in the strong-coupling
phase\cite{qd-us2}, to a description of which we now turn.

\subsection{The Effective Theory of the Strong-Coupling Phase}
\label{large-N-eff}

The strong-coupling phase occurs for $u_m<u_m^*$ for any $m$. We will
consider the simplest situation, when just one angular momentum
channel undergoes an instability. This is not an artificial
restriction for two reasons: Firstly, as seen in the RG treatment, the
instabilities in different channels are independent. Secondly, it is
highly unlikely that more than one Landau parameter is close to an
instability in a generic system. Furthermore, we will restrict our
attention to charge-channel instabilities in the spinful case, in
which the spin index is a passive spectator. This may turn out to be
the experimentally relevant type of instability; looking at the Landau
parameters\cite{kwon-ceperley} for the clean bulk 2DEG at $r_s=5$
(Eq. (\ref{fli-parameters})) we see that the $m=2$ charge channel is
the one closest to an instability. Instabilities in the spin channel
are extremely interesting, but in treating them one must necessarily
also account for the exchange coupling $J$. The resulting theory thus
depends on two parameters $(Z_m,J)$, and will be investigated in
detail in a future publication.

Thus, our model Landau
interaction function for this section is \beq u_m\cos{m(\t-\t')}
\eeq with no sum over $m$. All the thermodynamic properties of the
system can be obtained from the partition function $Z$, which can
be converted into a path integral in imaginary time by performing
the usual time-slicing according to \beq Z=Tr\bigg(e^{-\beta
H}\bigg)=Tr\lim_{\cN\to\infty}\prod\limits_{i}^{\cN}e^{-\beta
H/\cN} \eeq where $\beta=1/T$ is the inverse temperature.  One
then inserts a complete set of fermionic coherent states between
each factor of $e^{-\beta H/\cN}$, and writes the matrix elements
in terms of the coherent state labels. In the process, the
fermionic operators $c,c^{\dagger}$ get replaced by Grassmann
numbers, which we will represent by $\psi, \bar {\psi}$. The
partition function can now be written as \beq Z=\int D\psi
D{\bar\psi} e^{-S} \eeq where \beqr
S=&\int\limits_{-\beta/2}^{\beta/2} dt
\bigg(\sum\limits_{\a}{\bar\psi}_\a(t)({\partial\over\partial
t}+\ve_\a)\psi_\a(t) \nonumber\\
&-{|u_m|\Delta\over2}\sum\limits_{\bk,\bk'} \cos{m(\t-\t')}
n_{\bk}(t)n_{\bk'}(t)\bigg). \eeqr Here
$n_{\bk}(t)={\bar\psi}_{\bk}(t)\psi_{\bk}(t)$ and we have made
explicit  the fact that $u_m<0$.  We factorize the interaction
term in one time slice (of thickness $\delta t=\beta/\cN$):
\begin{eqnarray}
&\exp\left[\delta t{\Delta \over 2}\sum_{\bk \bk'}|u_m|\cos
(m\theta - m\theta')n_{\bk} n_{\bk'}\right]=\nonumber\\ &\int
d\sigmab \exp{-\delta t \Delta\left[g^2{|\sigmab|^2\over 2|u_m|}
+g\sum\limits_{\bk}n_\bk
(\sigma_1\cos{m\t}+\sigma_2\sin{m\t})\right]}\nonumber\\ &\!\!\!\!
=\int d\sigmab \exp \left[ -g^2\delta t \Delta{|\sigmab|^2\over
2|u_m|} -g\delta t\Delta\sum_{\a\b}\bar{\psi}_{\alpha}\sigmab
\cdot \bM_{\alpha \beta}\psi_{\beta}\right]
\label{hubb-strat}\nonumber
\end{eqnarray} where $ \sigmab =(\sigma^1\ ,
\sigma^2 )$  has two components,  as does $\bM$:
$(M^1,M^2)_{\a\b}=\sum_{\bk}\phi^{*}_{\alpha}(\bk
)\phi^{}_{\beta}(\bk )\left( \cos m \theta  \ , \sin m\theta
\right)$. Note that this factorization with a real action for
$\sigmab$ can be carried out only for $u_m<0$, which is indeed the
case here. The factors of $g$ are chosen for later convenience.

The fermionic action is now quadratic, and the fermions can be
integrated out to obtain an effective action for $\sigmab$ \beq
S_{eff}= -Tr \ln\left[ (i\omega -\varepsilon_\alpha )\bI
-g\Delta\sigmab \cdot \bM \right] +g^2\Delta\int dt
{|\sigmab|^2\over 2|u_m|} \label{effective} \eeq where $I$ is the
unit matrix and $\sigmab$ is now a function of imaginary time $t$.

To make further progress we expand $S_{eff}$ in powers of $\sigmab$,
which can be expressed graphically by the set of ``ring diagrams''
shown in Fig. 3.

\begin{figure}
\narrowtext \epsfxsize=2.4in\epsfysize=2.0in \hskip
0.3in\epsfbox{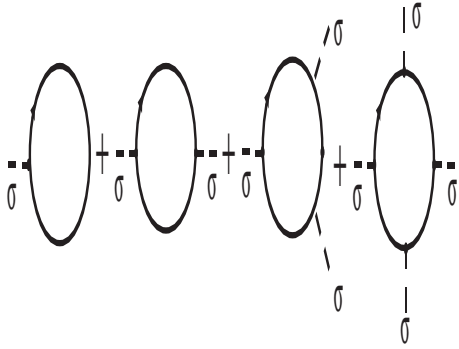} \vskip 0.15in \caption{The Feynman
diagrams in the graphical expansion of the Trace Log. Each diagram
has one fermion loop with different numbers of external $\sigmab$
legs.} \label{ring-diagrams}
\end{figure}

Evaluating the diagram with two external $\sigmab$ legs, we find the
following quadratic contribution from the $Tr\ln$ in the static limit
(when fluctuations of $\sigmab$ in imaginary time are ignored):
\begin{eqnarray} &g^2\Delta^2\int dt
\sum\limits_{\a\b}{N_F(\b)-N_F(\a)\over\varepsilon_\a-\varepsilon_\b}
\sum\limits_{\bk,\bk'}\phi_{\a}^*(\bk)\phi_\b(\bk)\phi_{\b}^*(\bk')\phi_\a(\bk')
\nonumber\\
&(\sigma_1\cos{m\theta}+\sigma_2\sin{m\theta})(\sigma_1\cos{m\theta'}+\sigma_2\sin{m\theta'})
\label{quadratic}\end{eqnarray} where $N_F(\a)$ is the Fermi
occupation of the single-particle state $\a$.

This quantity can be recognized as the one that appears in the
internal summation of the leading large-$N$ one loop bubble, Eq.
(\ref{diagram-leading}). As mentioned before (and shown in
Appendix A) this quantity is self-averaging; its disorder-average
dominates its fluctuations by $g$. We carry out this
self-averaging using RMT to obtain the average quadratic part of
the action for static $\sigmab$ \beq {\bar S}_0 =g^2 \int dt
{|\sigmab|^2\Delta\over 2}\left[{1\over |u_m| }-{\ln 2
}\right]\label{eff-action-s0} \eeq where the bar represents the
fact that we are considering only the (self-) disorder-averaged
part of $S_0$, and the $\ln 2=1/|u_m^*|$ arises from the summation
over energies in Eq. (\ref{quadratic}). For the spinful case this
number would be twice as large. It is clear that for
$|u_m|<|u_m^*|$, $\sigmab=0$ is a stable solution, while $\sigmab$
will break symmetry for $|u_m|>|u_m^*|$. In both cases let \beq r=
{1\over |u_m|}-{1\over |u_{m}^{*}|} \label{r}\eeq denote the
distance from criticality.

Will this symmetry-breaking, present in the ``bare'' theory,
survive  fluctuations? The answer to this hinges on the fact that
a factor $g^2$ appears at all higher loops, i.e., in front of the
entire self-averaged action for static $\sigmab$. The
$4^{th}$-order term suffices to clarify the issue. The generic
$4^{th}$-order term is shown in Figure 4, and is represented by the
following expression (all factors except for $g$ have been
suppressed): \beq g^4\sigma^4\sum\limits_{\a\b\g\d}
{M_{\a\b}M_{\b\g}M_{\g\d}M_{\d\a}\over \ve^3} \eeq where $\ve$ is
a generic energy denominator.
\begin{figure}
\narrowtext \epsfxsize=2.4in\epsfysize=2.0in \hskip
0.3in\epsfbox{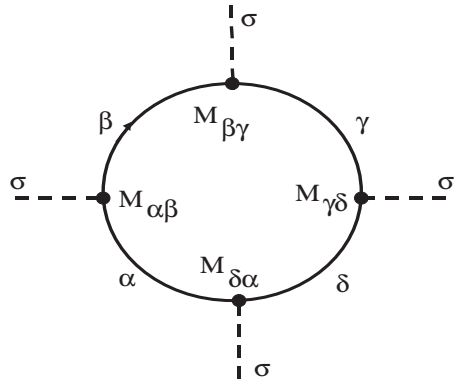
} \vskip 0.15in \caption{The Feynman diagram
corresponding to the fourth-order term in the effective action. }
\label{fourth-order}
\end{figure}

We now  replace the product of wavefunction sitting inside the
$M$'s by their averages, giving the usual arguments about
self-averaging. Just as the  average of four $\phi$'s had many
terms corresponding to different contractions of indices (Eqn.
(\ref{4pt1})),  there are many terms here as well. In the leading
contribution, the $\phi_{\b}$ from $M_{\a\b}$ is contracted with
the $\phi_{\b}^{*}$ from $M_{\b\gamma}$ and so on around the loop.
This forces the same $\bk$ index to flow around the loop,
generating the factor $\sum_{\bk} 1/g^4=1/g^3$.  There are 4
energy summations (going up to $g\Delta$) and three factors of
energy in the denominator (each typically of order $g\Delta$),
leading to an extra factor of $g$. The overall result is therefore
of order $g^2\sigma^4$.

It can be shown along very similar lines that the factor of $g^2$
can be extracted from the entire self-averaged effective action in
the static limit. Thus $g^2$ plays the role of $1/\hbar$ (or the
inverse temperature), and controls the size of fluctuations around
the saddle-point. This is what renders the large-$N$ approximation
realistic for ballistic dots with large dimensionless conductance.

Armed with the factor of $g^2$ multiplying the entire static
self-averaged effective action, one can see that the saddle-point
approximation becomes exact in the large-$g$ limit, and therefore
our conclusion that symmetry-breaking  takes place for
$|u_m|>|u_m^*|$ is in fact correct in this limit.

To see what the order parameter corresponds to, recall that in
terms of the original fermions
$\langle(\sigma_1,\sigma_2)\rangle=\sum_{\bk} \langle n_\bk\rangle
(\cos{m\theta},\sin{m\theta})$, where $\langle\rangle$  denotes a
quantum-mechanical (but not disorder-ensemble) averaging.  This
shows that some values of $\bk$ are preferentially occupied
compared to others. Therefore this can be identified as the
mesoscopic, disordered, version of the Pomeranchuk shape
transition of the Fermi surface\cite{pomeranchuk,agd}. There has
been a revival of interest in the bulk transition
recently\cite{varma,oganesyan}. Due to the disorder, $\bk$ is not
a good quantum number, and $n_{\bk}$ will suffer quantum
fluctuations, unlike the clean bulk case. The quantum-mechanical
average $\langle n_{\bk}\rangle$ is similarly not restricted to be
0 or 1. The Fermi ``surface'' will not be sharp, but will be
smeared out by the disorder. However, the angular average
corresponding to $\sigmab$ still provides an unambiguous measure
of symmetry-breaking.

\subsection{The Connection of Large-$N$ to the Hartree Approximation}
\label{large-N-HF}

The static large-$N$ saddle-point solution is the same as standard
Hartree mean field theory (the Fock term is down by $1/g$ because
of the form of the Fermi liquid interaction). This can be seen
very easily by noting that the saddle point satisfies the
self-consistency condition \beq {d\over d\sigmab}
\bigg(-Tr\ln{[(i\w-\ve_\a)\bI -g\Delta\sigmab \cdot \bM]}
+g^2\Delta\int dt {|\sigmab|^2\over 2|u_m|}\bigg)=0
\label{self-con1}\eeq

The above equation in fact minimizes the expression for $\beta$
times the free energy, or just the ground state energy as $T\to
0$. Note that it has two parts, one from the fermionic $Tr \ \ln$
and one from the $\sigmab$ field energy,\  $|\sigmab|^2\Delta\over
2|u_m| $.

Using the Feynman-Hellman theorem, which equates the $\sigmab$
derivative of the expectation value of the fermion hamiltonian to
the expectation value of the derivative of the fermion
hamiltonian,  we have \beqr &T{d\over d\sigmab}
\bigg(-Tr\ln[(i\w-\ve_\a)\bI -g\Delta\sigmab \cdot \bM ]\bigg)=\nonumber\\
&g\Delta\langle SD(\sigmab)|\sum_{\bk}n_\bk (\cos{m\t}\hi
+\sin{m\t}\hj)|SD(\sigmab)\rangle \eeqr where we have introduced
the Slater Determinantal state $|SD(\sigmab)\rangle$, which is
obtained by diagonalizing the Hamiltonian with a static $\sigmab$,
and filling up the lowest $g/2$ levels. This leads to the
self-consistency condition \beq u_m \langle
SD(\sigmab)|\sum_{\bk}n_\bk (\cos{m\t}\hi
+\sin{m\t}\hj)|SD(\sigmab)\rangle = g\sigmab \eeq

This is exactly the condition one would get by decoupling the
interaction in a Hartree approximation, which indicates that at
the saddle-point level, the large-$N$ theory is the same as the
Hartree approximation. However, the large-$N$ nature of the theory
allows one to justify the saddle-point, and to systematically
calculate corrections to it.

\section{Phase Diagram and Physical Signatures}
\label{phase-diagram}

We now turn to  qualitative and quantitative consequences of the
large-$g$ solution. Our discussion will use the  language of phase
transitions - collective  variables becoming gapless and order
parameters turning on at criticality - but there will be
inevitable modifications due the fact that we are primarily
interested in a mesocopic system with discrete energy levels.

 Let us begin by reiterating that  a genuine phase transition occurs
only in the limit  $g \to \infty$. This  is a mathematical
idealization that allows one to study and classify dots, but is
inaccessible in practice because  $\Delta \simeq 1/L^2, E_T
\simeq 1/L$, and their ratio $g\simeq L$. Thus  the limit $g\to
\infty$, corresponds $L \to \infty$, i.e., a dot of infinite size.
In such a system $E_T \to 0$ and everything we discuss (states
sensitive to boundary conditions and a phase transition that precedes
that of the clean bulk system) occurs within an energy window of zero
thickness and is thus invisible.  This is to be expected because in
this limit we are really describing an infinite ballistic
system. \footnote{This idealization was also implied in RG analysis
when it gave a zero of the $\b$ function at $u^*$.  In order to get
singular behavior out of a zero of a $\b$-function, we need to be able
to carry out the RG transformation an indefinite number of times. This
means $E_T$ must contain an infinite number of states. In a real dot
the flow ends when we come to within $\Delta$ of the Fermi
surface. Thus the critical state with $r=0$ is characterized by the
fact that $\sigmab$ has gap of order $\Delta$ as compared to the
generic value of $g\Delta$.}

We shall refer to the transition that occurs within $E_T$ in the
$g\to \infty $ limit as the mesoscopic Pomeranchuk  transition.
With the above words of explanation, the  reader should not have
any trouble reconciling the juxtaposition of the words mesoscopic
and phase transition in one phrase.

{While the $g\to\infty$ quantum dot is not directly accessible, we are
interested in it because several nontrivial features of the phase
transition it undergoes manifest themselves even for $1/g >0$.} This
result is based on recent investigations into quantum phase
transitions, which show that while the $T=0$ transition (as a function
of some coupling) occurs at an isolated point that is also physically
inaccessible, it controls the physical behavior in a fan-shaped
quantum critical regime\cite{critical-fan} in the temperature-coupling
constant plane. In our problem $1/g$ plays the role of $T$ since $g^2$
multiplies the action. The situation is depicted in Figure 4. The
dotted lines bounding the quantum critical regime indicate the sharp
crossover which replaces the transition.  There are however some
differences from the usual quantum critical descriptions because $g$
not only appears outside the action and plays the role of inverse
temperature, but also appears within $S_{eff}$ via subleading,
disorder-realization-specific terms which are especially important in
the strong-coupling phase.
\begin{figure}
\narrowtext
\epsfxsize=2.4in\epsfysize=2.0in
\hskip 0.3in\epsfbox{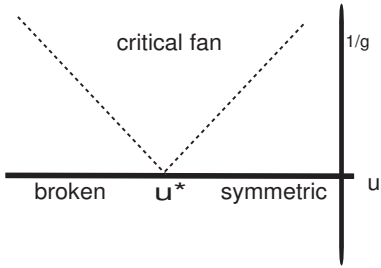}
\vskip 0.15in
\caption{The generic phase diagram for a second-order quantum phase transition.
The horizontal axis represents the coupling constant which can be
tuned to take one across the transition. The vertical axis is usually
the temperature in bulk quantum systems, but is $1/g$ here, since in
our system one of the roles played by $g$ is that of the inverse
temperature.}
\label{evschi}
\end{figure}

In this section we will explore the three regimes: weak-coupling,
strong-coupling, and quantum-critical as well as crossovers between
them.  The physics of the weak-coupling regime is controlled by the
Universal Hamiltonian. However, we will point out that even in a
system belonging to this regime one can access the physics of the
quantum-critical regime by making a finite frequency measurement. Thus
even dots in this region can be used to confirm the existence of the
phase transition we have found at $u^*$.

We first turn to the strong-coupling and quantum-critical regimes
 which are new to our work.

\subsection{The Strong-Coupling Regime}
\label{strong-coupling-phase}

As shown by Eq. (\ref{eff-action-s0}) $\sigmab$ acquires a nonzero
expectation value if $u_m<u_m^*$ or $r<0$. Since $\sigmab$ can be
thought of as a two-dimensional vector order parameter, and the
self-averaging part of $S_{eff}$ has rotational symmetry, one
expects a Mexican Hat (a circle of exactly degenerate minima
$|\sigmab|=\sigma_0$). For $g=\infty$ the system can still break
symmetry and pick a particular value of $\sigmab$, since there are
no kinetic term for $\sigmab$ in that limit (see Eq.
(\ref{eff-action-dynamic}) below). However, for $1/g\ne0$ one may
expect that $\sigmab$ will delocalize immediately around the
Mexican Hat, restoring symmetry, since $1/g^2$ plays the role of
$T$ or $\hbar$ (depending on how one wants to interpret the path
integral). In other words one expects the zero-dimensional analog
of the Mermin-Wagner theorem\cite{mermin-wagner}. In terms of the
parameterization of $\sigmab$ in cylindrical coordinates \beq
\sigmab=(\sigma,\chi) \eeq we expect the $\sigmab$ variable to
uniformly populate all values of the angle $\chi$.

This does not actually happen, thanks to disorder-realization-specific
terms in $S_{eff}$, which are subleading in $1/g$ to the terms
considered so far and explicitly break rotational symmetry in
$\sigmab$ space. There are many sources for such $1/g$ terms.  The
coefficient of the quadratic term coming from the $Tr\ln$, which has a
leading average part which produces the $\ln{2}$, also has
sample-to-sample fluctuations which are down by $1/g$ (as shown in
Appendix A). In fact, a term {\it first-order} in $\sigmab$ is also
present for a system undergoing an instability in an even angular
momentum channel $m$. These subleading terms render the valley of the
Mexican Hat nondegenerate in the case of even $m$, and leave behind a
two-fold degeneracy in the case of $m$ odd.

Will these sample specific terms localize $\sigmab$ in the angular
direction? To resolve this issue we need to understand the low-energy
dynamics of $\sigmab$. To this end we compute the quadratic part of
the effective action for slow fluctuations of $\sigmab$.  We find \beq
{\bar S}_0 =g^2 \int dt {|\sigmab|^2\Delta\over 2}\left[{1\over |u_m|
}-{1\over |u_m^*| }\right]+ g\int {d\w\over2\pi} |\sigmab(\w)|^2
f(\w)\label{eff-action-dynamic} \eeq where the bar represents the fact
that we are considering only the (self-) disorder-averaged part of
$S_0$. The frequency dependent function $f(\w)$ is \beqr
f(\w)=&{1\over4} \bigg( 2\w\tan^{-1}{\w\over\Delta} +
2\w\tan^{-1}{\w\over\Lambda}-4\w\tan^{-1}{2\w\over\Lambda}\nonumber\\
&+\Lambda\ln\left[{(4\w^2+\Lambda^2)\over(\w^2+\Lambda^2)}\right]-\Delta\ln\left[1+{\w^2\over\Delta^2}\right]\bigg)
\label{freq-dependence-s0}\eeqr Let us note that

 \beqr f(\w)&=&\w^2/4\Delta \ \ \ \ \ \ \w\ll \Delta \label{ham} \\
 &=& \pi|\w|/4 \ \ \ \ \ \ \ \ \ \Delta\ll\omega\ll E_T \label{diff}
 \eeqr Thus we expect Hamiltonian dynamics at very low energies, with
 a non-universal sample specific dependance on the lower cutoff
 $\Delta$. At high energies we have overdamped dynamics, representing
 the effect of Landau damping on the $\sigmab$ variable. This latter
 behavior is universal and self-averaging.\footnote{The reader might
 ask what permits us to keep some terms which are $1/g$ down compared
 to the self-averaged static part of $S_{eff}$, and discard
 others. The answer is that we keep only the lowest order terms of
 every type. Thus, the realization-specific terms which break the
 rotational symmetry of the self-averaged part of $S_{eff}$ are kept,
 while $1/g$ corrections to the self-averaged part are
 discarded. Similarly, though the kinetic term of
 Eq. (\ref{eff-action-dynamic}) is order $1/g$ down compared to the
 self-averaged part, it must be kept since this is the lowest order
 dynamics.}

At very low energies described by hamiltonian dynamics  Eqs.
(\ref{ham},)  the  momentum conjugate to  $\sigmab$ is \beq
 {\bf
p}_{\sigmab}=g{\dot {\sigmab}}/2\Delta .\eeq
Only the angular
dynamics is important in the symmetry-broken phase since the
radial fluctuations of $\sigmab$ are suppressed by the Mexican Hat
potential of order $g^2\Delta$. The Hamiltonian for small
oscillations around the global minimum of the explicit
symmetry-broken effective action looks like

\beq H_{osc}=\Delta{{\hat L}^2\over g}+g\Delta V(\chi)
\label{osham}\eeq where we have denoted the angular part of the
canonical momentum as ${\hat L}$. The spread of the ground state
wave function (of this oscillator)  can be estimated as
$\delta\chi\simeq 1/\sqrt{g}$. Thus at large $g$ the support of
$\sigmab$ is localized to the minimum of the potential energy
term. Our analysis is internally consistent: The form of $f(\omega
)$ we used to  obtain oscillator description (Eqn. (\ref{ham})) is
valid out to $\omega \simeq \Delta$, the oscillator ground state
energy. \footnote{ The low frequency approximation $f(\omega )
\simeq \omega^2/(4\Delta )$ tracks the exact expression to with
ten percent in this range.}

Thus   symmetry-breaking will occur in the strong-coupling regime
for large enough $g$ when disorder-realization-specific terms are
included. The result  is a consequence  of spontaneous
symmetry-breaking (requiring strong interactions) and  explicit
symmetry-breaking (from disorder).

These considerations apply uniformly for $m$ even and odd. However,
there are significant differences in other features and we will treat
them separately.

\subsubsection{The Case of $m$ Even}

Let us see how to find the effective potential with the
realization-specific corrections more explicitly. We need to find
the ``potential landscape'' of $\sigmab$, which we define for
$t$-independent $\sigmab$ as \beq
V_{eff}(\sigmab)=g^2\Delta{\sigmab^2\over2|u_m|}-{1\over\beta}Tr\ln[(i\w-\ve_\a)\delta_{\a\b}-g\Delta\sigmab\cdot\bM_{\a\b}]
\label{effective-pot}\eeq where $\beta=1/T$ is the inverse
temperature not to be confused with the label for  noninteracting
disorder eigenstates.  We thus need to map out the $Tr\ln$ for
time-independent $\sigmab$. In the limit of zero temperature, the
quantity $Tr\ln(i\w-H_{F,\sigmab})/\beta$ is just
$-\cE_F(\sigmab)$, where the fermionic Hamiltonian is \beq
H_{F,\sigmab}=\sum\limits_{\a} \ve_\a c^{\dagger}_\a c_\a
-g\Delta\sigmab\cdot\sum\limits_{\a\b}\bM_{\a\b}c^{\dagger}_\a
c_\b \label{Hsigmab}\eeq and $\cE_F(\sigmab)$ is its ground state
energy, obtained by diagonalizing the hamiltonian $H_{F,\sigmab}$
and filling up the lowest states with the requisite number of
fermions. Let us note that for large $g$, since fluctuations in
$\sigmab$ are small, we can approximate the {\it total energy of
the interacting system with $N$ particles} $\cE_N$ as the minimum
of the effective potential of Eq. (\ref{effective-pot}), \beq
\cE_N= Min_{\sigmab}\bigg(\cE_F(\sigmab)+{g^2\Delta \sigmab^2
\over 2|u_m|} \bigg) \label{e-interacting}\eeq where
$Min_{\sigmab}$ indicates that the argument should be minimized
over all $\sigmab$. Note that the value of $\sigmab$ at the
minimum depends on the realization as well as the number of
particles $N$.

For small
$\sigmab$ we can find $\cE_F$ by standard perturbation theory.  To
first order we find the following energy correction \beqr
\cE_F^{(1)}(\sigma,\chi)=&-g\Delta\sigmab\cdot\sum\limits_{\a occ}
\bM_{\a\a}\\ =&-g\Delta\sigma\sum\limits_{\bk,\a occ}
\phi^*_{\a}(\bk)\phi_\a(\bk)\cos(m\t-\chi) \eeqr where the sum is
over occupied states only, and $\sigma,\ \chi$ are the magnitude
and angle of $\sigmab$ respectively. From the fact that
$\phi^*_\a(\bk)=\phi_\a(-\bk)$ it follows that \beq
\bM_{\a\b}=(-1)^m\bM^*_{\b\a} \label{M-identity}\eeq a specific
instance of which is \beq \bM_{\a\a}=(-1)^m\bM_{\a\a} \eeq This in
turn shows that $\bM_{\a\a}$ must vanish for odd $m$. However, for
even $m$ there is no reason for this to vanish for a particular
disorder realization (though its disorder-average does vanish).
One can also see that this term produces a dependence on the angle
$\chi$ of $\sigmab$, which breaks the degeneracy of states in the
Mexican Hat. To get an idea of the magnitude of this
realization-dependent term, we square it and average over
disorder. It is easy to verify that the disorder ensemble average
(denoted by $\ll\gg$)is given by

\beq \ll (\cE_F^{(1)})^2 \gg \simeq g^2 \Delta^2\eeq meaning that
the typical size of this correction is of order $g$, one power of
$g$ down from the self-averaging part of $\cE$.

One can also analytically get the statistical correlations in the
shape of the first-order contribution to the potential by
computing

\beq \ll \cE_F^{(1)}(\chi_1)\cE_F^{(1)}(\chi_2) \gg \simeq g^2\Delta^2
\cos(\chi_1-\chi_2) \eeq

The picture we get for even $m$
is then the following: Each order in perturbation theory
contributes a fairly smooth (on average) realization-dependent
term which depends on $\chi$, thus breaking the symmetry of the
$g=\infty$ Mexican Hat. If the system is not too deep in the
strong-coupling phase we can expect that only a few orders of
perturbation theory contribute, and that the angle-dependent
potential in the Mexican Hat is fairly smooth, and generically has
a single minimum. The most important effect of the
realization-dependent terms is to allow for symmetry-breaking even
for finite $g$. {\it Therefore, both disorder and interactions are
crucial to this symmetry-breaking: Interactions produce the
Mexican Hat potential causing  spontaneous symmetry-breaking at
$g=\infty$,  while explicit sample specific symmetry breaking
terms  prevent the restoration of symmetry by quantum
fluctuations even at finite $g$}.

\begin{figure}
\narrowtext
\epsfxsize=2.4in\epsfysize=2.0in
\hskip 0.3in\epsfbox{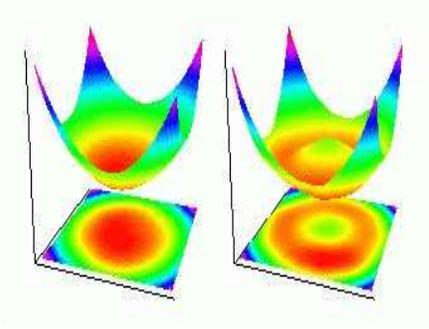}
\vskip 0.15in
\caption{The effective potential as a function of $\sigmab$ for  $m=2$,
with $g=20$ and $u_m=-1.2$ (left panel) and $u_m=-1.7$ (right
panel). The symmetry-breaking in the second case is clear. }
\label{mexican-hat-m-2}
\end{figure}

At $T=0$ one can simply evaluate the fermionic ground state energy
$\cE_F(\sigmab)$ corresponding to each static $\sigmab$
numerically. We show the resulting effective potential of Eq.
(\ref{effective-pot}) in Fig. 6 for the illustrative case $m=2$.
As can be seen, there is an approximate Mexican Hat circle of
degenerate minima. In Fig. 7 we zoom in on this circle to see that
the degeneracy is broken by sample-specific disorder and that the
fluctuations are indeed  of order $1/g$ relative to the
self-averaging term.   Clearly, the numerical work corroborates
the analytical picture developed above.

\begin{figure}
\narrowtext \epsfxsize=2.4in\epsfysize=2.0in \hskip
0.3in\epsfbox{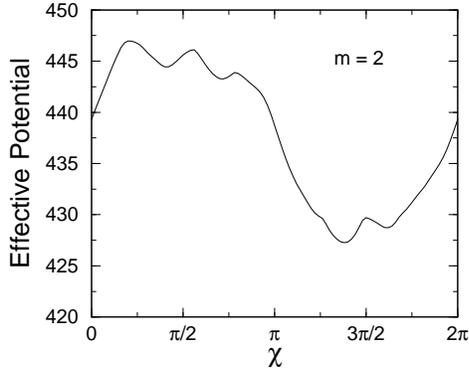} \vskip 0.15in \caption{The effective
potential in the Mexican Hat in units of $\Delta$ plotted as a
function of the angle $\chi$ for $m=2$, at $g=20$ and $u_m=1.7$.
Note the single nondegenerate minimum and the scale of sample
specific disorder, which is of order $1/g$ relative to the
average. } \label{evschi-m-2}
\end{figure}

One can now ask how various measurable quantities behave in the
strong-coupling phase. There are three such quantities of
interest. \begin{itemize} \item The addition spectrum, which measures
the difference in ground state energies
$\Delta_1(N)=\cE_{N+1}-\cE_{N}$ where the total energy of the system
with $N$ particles $\cE_N$ is the sum of the fermionic energy and the
energy of the bosonic field $\sigmab$ (Eq. (\ref{e-interacting})). We
have not included the charging energy $U_0 N^2/2$.  The Coulomb
Blockade peak spacing is then given by
$\Delta_2(N)=\Delta_1(N+1)-\Delta_1(N)$.
\item The peak-height distribution.
\item  The persistent current
in response to an external flux.
\end{itemize}

 Since at any particular $\sigmab$ the
electronic excitations are controlled by a single-particle
Hamiltonian, and $\sigmab$ is well-localized, one might naively
imagine that the addition spectrum is the same as the
level-spacing spectrum, which is the Wigner-Dyson distribution
with an enhanced $\Delta$. However, this ignores possible shifts
of $\sigmab$ upon adding a particle. Most of the time the value of
$\sigmab$ shifts only a little upon adding a particle. However,
every now and then a minimum which was far away in $\sigmab$-space
becomes lower upon adding a particle. One can compute the peak
spacing from $\Delta_2(N)=\Delta_1(N+1)-\Delta_1(N)$. In Figure 8
we show the distribution of $\Delta_2$ calculated numerically by
the method explained above Eq. (\ref{Hsigmab}). As can be seen,
negative values of $\Delta_2$ are possible because of the rare
events in which $\sigmab$ changes discontinuously upon adding a
particle, which makes the distribution broader and more symmetric
than the level-spacing distribution. (Note once again that the
negative answer results only upon subtracting out the charging
energy $U_0$.) In contrast negative values of $\Delta_2$ are
impossible in a noninteracting model.

\begin{figure}
\narrowtext
\epsfxsize=2.4in\epsfysize=2.0in
\hskip 0.3in\epsfbox{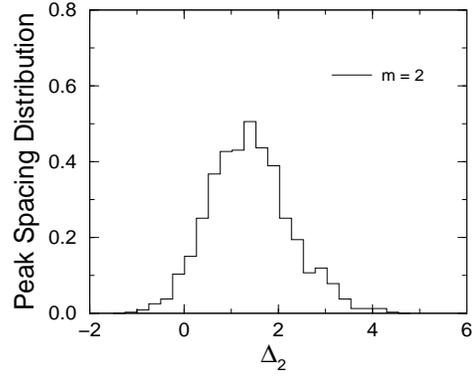}
\vskip 0.15in
\caption{The peak-spacing distribution for $m=2$, at $g=20$ and $u_m=1.7$.
Note the incidence of negative values of the peak-spacing, which is
impossible in a noninteracting model. }
\label{delta2-m-2}
\end{figure}

Now consider the peak heights. In the noninteracting case, the
distribution has been computed in RMT by Jalabert, Stone, and
Alhassid\cite{peak-height-th}, and measured experimentally in the
weak-coupling regime at very low
temperatures\cite{peak-height-expt}. There are characteristic
differences\cite{peak-height-th} between the peak-height distributions
for the GOE (time-reversal unbroken) and GUE( time-reversal broken)
which have been measured\cite{peak-height-expt}, and will play a role
in the following subsection. There are discrepancies between theory
and experiment at higher temperatures\cite{peak-height-expt2} which we
will address qualitatively in a later subsection. As long as $\sigmab$
changes by small amounts when particles are added the strongly-coupled
system behaves much like the noninteracting system. However, when a
large change of $\sigmab$ is involved, all the electrons must change
their state to adjust to the new $\sigmab$. This means the transition
matrix element the fermion creation operator between the two ground
states will be small, and we expect the peak height to be considerably
smaller.\footnote{Changing $\sigmab$ is the same as changing a
parameter in a noninteracting random Hamiltonian, a theme which has
been explored in the literature\cite{parametric}. Borrowing the result
for the overlap of determinants\cite{orthogonality}, for ground stated
differing by $\delta\sigmab$ we expect the peak height to scale as
$e^{-g\ln{g}(\delta\sigmab)^2}$.}This effect produces a correlation
between smaller than average {\it peak spacing} and small {\it peak
height}.

Finally, let us briefly consider the response of the system to an
external flux, leaving the details to Part II\cite{partII} (a short
report of this work can be found in ref. \cite{short}). This response
is none other than the persistent current \beq I_{pers}=-{\partial
F\over\partial\phi} \eeq where $F$ is the Free energy (the ground
state energy at $T=0$).  This response is primarily orbital in $GaAs$
due to the tiny coupling of the external flux to the spins.

\begin{figure}
\narrowtext
\epsfxsize=2.4in\epsfysize=2.0in
\hskip 0.3in\epsfbox{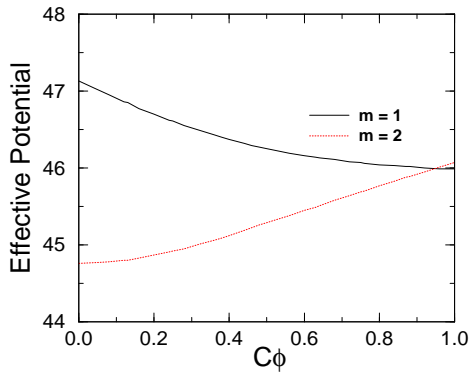}
\vskip 0.15in
\caption{The ensemble-averaged ground state energy as a function of
external flux for $m=1$ (solid line) and $m=2$ (dashed line), at
$g=20$ and $u_m=1.7$. Note the fact that the ground state energy
increases for $m=2$, which thus shows diamagnetic behavior, while
$m=1$ shows paramagnetic behavior. }
\label{egs-vs-phi-m-12}
\end{figure}

Figure 9 shows the ground state energy as a function of external
flux averaged over many disorder realizations. The noteworthy
point is that the average persistent current is diamagnetic. To
see why this is special, a brief digression into what is known
about persistent currents is necessary. In a mesoscopic ring
penetrated by a flux, the ground state energy has to be periodic
in the flux, since an integer number of flux quanta can be gauged
away. So \beq I_{pers}(\phi)=-{\partial
F\over\partial\phi}=I_1\sin(2\pi\phi/\phi_0)+I_2\sin(4\pi\phi/\phi_0)
+\cdots \eeq where $\phi_0=h/e$ is the flux quantum.

The noninteracting problem is relatively
well-understood\cite{mesoscopics-review,persist-nonint-th}. Only
the even moments $I_{2n}$ survive disorder-averaging (this result
holds for the interacting case as well). The {\it typical},
fluctuating values of the Fourier coefficients are (for small $n$)
$I_{n,typ}\approx {E_T/\phi_0}$ while the average is $ \langle
I_{2n}\rangle\approx {\Delta/\phi_0}$.

Interactions, when included in renormalized first-order
perturbation theory\cite{persist-int-th}, produce \beq \langle
I_{2}\rangle\approx \mu^*{E_T\over\phi_0}
\label{ave-current-int}\eeq where $\mu^*$ (typically $<1$) is the
dimensionless Cooper-channel interaction at low energies. The
conclusion is that interactions enhance the average persistent
current, but that if $\mu^*>0$ $\langle I_2\rangle$ should be
paramagnetic, while if $\mu^*<0$ it should be diamagnetic. Much
numerical work has followed since\cite{persist-numerics}, mostly
on one-dimensional rings, confirming that interactions do indeed
enhance $\langle I_2\rangle$ in the spinful case.

Experiments\cite{persist-expt} reveal striking discrepancies with
the above predictions. The predicted value\cite{persist-int-th},
while of the right order-of-magnitude, is still smaller than
experiment. More disturbingly, {\it the sign is inconsistent}.
Even materials that show no sign of superconductivity (implying
that $\mu^*>0$) show\cite{persist-expt} a diamagnetic $\langle
I_2\rangle$. Many other explanations (e.g. ref.
\cite{kravtsov,far-levels}) have been proposed to account for
these observations, but questions about the sign and the magnitude
of persistent currents in non-superconducting materials remain
open\cite{persist-puzzle}.

In this context it is striking to obtain, as we do for $m$ even, a
diamagnetic persistent current in a model without
superconductivity.

All the signatures described above (expect for the diamagnetic
persistent current) also hold for the case of $m$ odd. However,
there are further interesting effects which we describe below.

\subsubsection{The Case of $m$ Odd}
\label{oddm}

Let us now turn to odd $m$, where the first-order contribution in
$\sigmab$ to the effective potential is missing. In fact, one can
show a much more general result, that the ground state energy to
all orders $\cE(\chi)$ at a particular angle $\chi$ with $\sigma=|\sigmab|$ fixed satisfies

\beq \cE(\chi)=\cE(\chi+\pi) \eeq

To show this result we write the second-quantized  fermionic Hamiltonian for
the illustrative case of $m=1$ as

\beq H_F(\chi)=\sum\limits_{\a} \ve_\a c^{\dagger}_\a c_\a
-g\Delta\sigma\sum\limits_{\a\b,\bk}c^{\dagger}_\a c_\b
\phi^*_\a(\bk)\phi_\b(\bk)\cos(\t-\chi) \label{Hchi}\eeq and its
first-quantized version: \beq h_{\a\b}= \delta_{\a \b}\ve_\a
-g\Delta\sigma \sum\limits_{\bk}
\phi^*_\a(\bk)\phi_\b(\bk)\cos(\t-\chi). \eeq

Using Eqs. (\ref{goe-k-minus-k-connection},\ref{M-identity}) along
with $\t(-\bk)=\t(\bk)\pm\pi$ one can easily show that

\beq h^*(\chi)=h(\chi+\pi) \eeq

Since the Hamiltonian is hermitian at every $\chi$ its eigenvalues
are real, and thus

\beqr h(\chi)\Phi_n(\chi)=\e_n(\chi)\Phi_n(\chi)\nonumber\\
\Rightarrow
h^*(\chi)\Phi_n^*(\chi)=\e_n(\chi)\Phi_n^*(\chi)\nonumber\\
\Rightarrow h(\chi+\pi)\Phi_n^*(\chi)=\e_n(\chi)\Phi_n^*(\chi)
\label{Hchi-Phi}\eeqr

which shows that all the single-particle energies of $h(\chi)$ are
shared by $h(\chi+\pi)$, so naturally the ground state energies
are the same as well. This implies that the energy as a function
of angle $\chi$ generically has two exactly degenerate minima. The
complex conjugate relationship between the single-particle
wavefunctions shows that the two minima are connected by a
time-reversal transformation $\cT$. This is easy to understand for
$m=1$. In the bulk an $m=1$ distortion is just a shift of the
Fermi surface. Since the underlying Hamiltonian is time-reversal
symmetric, any distortion should have the same energy as the
time-reversed distortion.

It is also important to understand that an $m=1$ distortion in the
quantum dot does not represent a current flowing in a particular
direction, as it would in the bulk: This is impossible in the steady
state in a closed dot. Rather, the $m=1$ distortion describes a
persistent circulating current within the dot.  One can understand
this by referring to Eq. (\ref{defn-of-k}), which shows that the state
we label by $\bk$ in the dot has a vanishing wavefunction at the
boundary (assuming a hard wall boundary condition for simplicity), and
therefore the current carried by it vanishes at the boundary of the
dot.

Once again, one can numerically corroborate all these deductions, as
shown in Figs. 10 and 11, which show the approximate Mexican Hat and
the double-degeneracy in it for the illustrative case of $m=1$.
\begin{figure}
\narrowtext
\epsfxsize=2.4in\epsfysize=2.0in
\hskip 0.3in\epsfbox{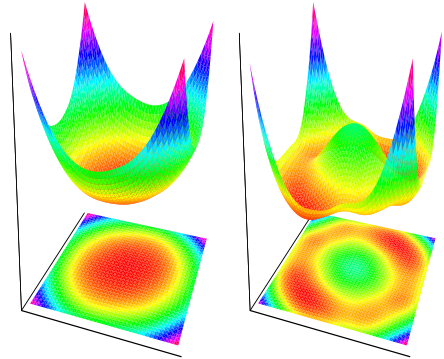}
\vskip 0.15in
\caption{The effective potential as a function of $\sigmab$ for  $m=1$,
with  $g=20$ and $u_m=-1.2$ (left panel) and $u_m=-1.7$ (right panel). }
\label{mexican-hat-m-1}
\end{figure}

\begin{figure}
\narrowtext \epsfxsize=2.4in\epsfysize=1.6in \hskip
0.3in\epsfbox{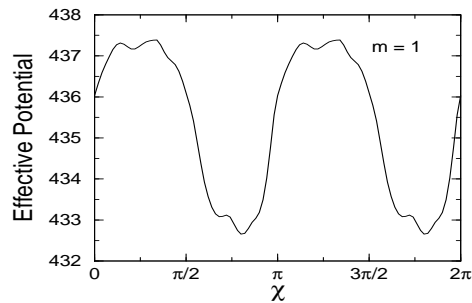} \vskip 0.15in \caption{The effective
potential in the Mexican Hat in units of $\Delta$ plotted as a
function of the angle $\chi$ for the case $m=1$, at $g=20$ and
$u_m=1.7$. Note the exact two-fold degeneracy of the effective
potential minima, and the scale of sample specific disorder, which
is of order $1/g$ relative to the average.  } \label{evschi-m-1}
\end{figure}

Similarly, the peak-spacing distribution can also be numerically
obtained, as shown in Fig. 12.
\begin{figure}
\narrowtext
\epsfxsize=2.4in\epsfysize=2.0in
\hskip 0.3in\epsfbox{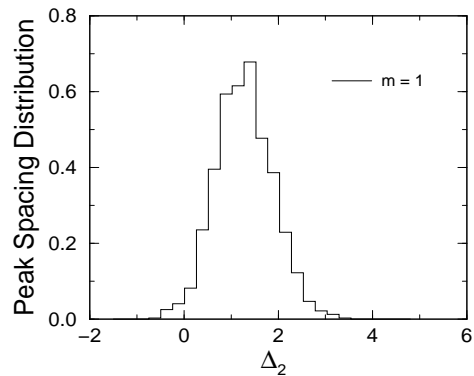}
\vskip 0.15in
\caption{The peak-spacing distribution for $m=1$, at
$g=20$ and $u_m=1.7$.  Note the incidence of negative values of the
peak-spacing, which is impossible in a noninteracting model. }
\label{delta2-m-1}
\end{figure}

Thus, the $O(2)$ symmetry of the Mexican Hat in the $g=\infty$
limit is broken down to a two-fold degeneracy for any finite $g$.
This leads to a very interesting physics which depends sensitively
on the coupling to the leads.

\begin{center}
\underline{Isolated Dot} \vskip0.1in
\end{center}

The isolated dot has two exactly degenerate minima in the
effective potential. However, the low-energy dynamics of $\sigmab$
will induce a ``tunneling'' term between the two minima. Recall
that the ``kinetic'' part of the effective action at very low
energies was approximately \beq S_{eff,kin}=g\int d\tau
\bigg({d\sigmab\over d\tau}\bigg)^2 \eeq The tunneling amplitude
typically goes as \beq A=e^{-\int dx \sqrt{2mV(x)}} \eeq where the
integral is over the classically forbidden region. In this problem
the mass, read off from the kinetic term, is of order $g$, and so
is the potential in the angular direction. Thus the tunneling
amplitude must be of order $\Delta e^{-g}$.

Thus the true eigenstates of the system will be the symmetric
superposition, which is the ground state, and the antisymmetric
superposition which is the excited state. The splitting between
them will be of order $e^{-g}\Delta$.The ground state at $T=0$
will not have any symmetry-breaking.

 However this state of affairs describes a very narrow low  temperature range: For
$\Delta\gg T\gg e^{-g}\Delta$ the system will be in an
(essentially)  equal incoherent mixture of the symmetric and
antisymmetric states, or equivalently, by  change of basis, in an
equal incoherent mixture of the two minima of the effective
potential. Each of these minima  has $\cT$-broken dynamics and GUE
statistics\cite{peak-height-th,peak-height-expt} {\it even in the
absence of an external magnetic field}. Since the two minima have
complex conjugate wavefunctions, the incoherent average will wipe
out the  the persistent currents which change sign under complex
conjugation,  but not the GUE peak-height
statistics\cite{peak-height-th} which are immune to complex
conjugation.

Another way, besides increasing  $T$ to experimentally observe
this degeneracy is to explicitly break time-reversal symmetry by
introducing a small external magnetic flux $\phi$. The response of
the system to such a flux will be linear instead of quadratic, as
in the time-reversal invariant cases. here are some specifics.

The  flux is introduced within the theory by replacing the
non-interacting part of the Hamiltonian by one drawn from an
ensemble of crossover Hamiltonians\cite{RMT}. Including the
external flux, the first quantized Hamiltonian at a particular
angle $\chi$ is (from Eq. (\ref{Hchi})) \beq h=h(\chi)+{C\
\phi\over\phi_0} h_{A} \label{hcross}\eeq where $C$ is a constant 
of order unity that depends on the shape of the dot and the nature of
boundary scattering\cite{univ-ham}, and $h_A$ is an appropriately
normalized, random, pure imaginary, antisymmetric matrix in any
orthogonal basis\cite{RMT} with typical matrix elements of size
$\Delta\sqrt{g}$. It is easy to show that the two minima of $V_{eff}$
move in opposite directions to first order in $\phi$. It is convenient
to use the exact noninteracting eigenstates as a preferred orthogonal
basis.  Using the notation of Eq. (\ref{Hchi-Phi}), the change in
energy to first order is \beq \delta \cE_F(\chi)=\sum\limits_{n\ occ}
\sum\limits_{\a\b} (\Phi_{n,\a}(\chi))^*
h_{A,\a\b}\Phi_{n,\b}(\chi) \eeq where $\Phi_{n,\a}(\chi)) $ is
the projection of the $n^{th}$ exact eigenfunction of Eq.
(\ref{hcross}) on the noninteracting disorder  eigenvector $\a$.
Noting from Eq. (\ref{Hchi-Phi}) that
$(\Phi_{n,\a}(\chi))^*=\Phi_{n,\a}(\chi+\pi)$, and using the
antisymmetry of $h_A$ in the basis of exact noninteracting
eigenstates, it follows that \beq \delta \cE_F(\chi)=-\delta
\cE_F(\chi+\pi) \eeq

One can estimate the magnitude of $\delta\cE_F$ by squaring and
averaging over the ensemble to be $\delta \cE_F\simeq (g\Delta)
\phi/\phi_0$. When this energy exceeds the splitting between the
symmetric and antisymmetric states, the system will localize
$\sigmab$ in the lower minimum of $V_{eff}$, thus breaking
time-reversal symmetry completely. The flux needed to achieve this
at $T=0$ is known as the crossover flux \beq \phi_{cross}\simeq
\phi_0 e^{-g}/g \eeq which is exponentially small in $g$. The
effective schematic hamiltonian in the subspace of the two minima,
labeled $+$ and $-$ is of the form \beq \Delta \left(
\begin{array}{cc} bg & \ e^{-g}
\\  e^{-g} & -bg
\end{array} \right)
\eeq where $b$ denotes the external magnetic field and $\Delta
e^{-g}$ the tunneling amplitude between the  two minima related by
time-reversal.

 This response to an applied flux should be contrasted with the
noninteracting case, when the crossover flux (for single-particle
properties) is of the order of $\phi_0/\sqrt{g}$. The persistent
current in the $m$ odd case is paramagnetic\cite{short}, as can be
seen from Fig. (\ref{egs-vs-phi-m-12}). The ground state energy
decreases linearly for small $\phi$, a consequence of the first order
contribution.

The response of the time-reversal broken system  manifests itself
in the shift in the zero-bias conductance peak position as a
function of external flux. Since the peak position reflects the
difference $\Delta_1=\cE_{N+1}-\cE_N$, and all the single-particle
energies change {\it linearly} with flux (for
$\phi_0/\sqrt{g}\gg\phi\gg\phi_{cross}$), the shift in the peak
position must also be linear in the external flux. This contrasts
with the noninteracting case where the shift is quadratic in the
external flux for $\phi_0/\sqrt{g}\gg\phi$. The numerical results
are shown in Fig 13.

\begin{figure}
\narrowtext
\epsfxsize=2.4in\epsfysize=2.0in
\hskip 0.3in\epsfbox{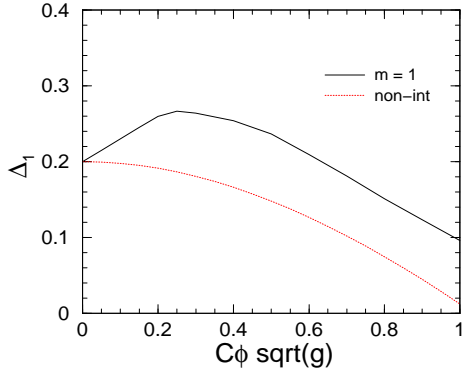}
\vskip 0.15in
\caption{The peak position $\Delta_1=\cE_{N+1}-\cE_{N}$  as a
function of external flux in the range $0<C\phi<\phi_0/\sqrt{g}$ for
$m=1$, at $g=50$ and $u_m=1.7$. Note the fact that the peak position
is linear in the flux for small values of flux in contrast to the
noninteracting result (dashed line) which is quadratic. }
\label{delta1-vs-phi-m-1}
\end{figure}

For $\Delta\gg T\gg e^{-g}\Delta$ one can carry out a
linear-response magnetization measurement, about which more will
be said in Part II\cite{partII}. The two degenerate states will lead to a $1/T$
Curie-like linear response.  Let us now turn to what happens when
the quantum dot is weakly coupled to the leads.

\begin{center}
\underline{Dot Coupled to Leads: Mapping to} \underline{the
Caldeira-Leggett and Kondo Problems for $m$ odd} \vskip 0.1in
\end{center}

We have already seen that at very low energies the dynamics of
$\sigmab$ in an isolated dot becomes Hamiltonian, due to the fact
that there are no particle-hole excitations below a lowest energy.
This changes if the dot is coupled to leads, since the dot levels
will hybridize with the continuum of levels in the leads. This
leads to dissipative dynamics for $\sigmab$ at arbitrarily low
energies. Since in the case of $m$ odd we know that there are two
degenerate minima of the effective potential, we are naturally led
to the Caldeira-Leggett problem\cite{CL}, which concerns a
particle in a degenerate double-well potential interacting with a
``bath'' of harmonic oscillators which model the dissipation.
Applying the solution\cite{chakravarty,bray-moore} of the
Caldeira-Leggett model to our problem, {\em we can predict that
there will be two phases depending on the strength of the coupling
to the leads: One with spontaneously broken time-reversal symmetry
where the $\sigmab$ field is trapped in one minimum, and one where
it can successfully tunnel back and forth between the two minima
and time-reversal symmetry is unbroken.}

Let us recall some of the principal results of  Caldeira and
Leggett\cite{CL}. Their motivation was to study systems with a macroscopic
quantum mechanical degree of freedom (such as the flux through a
superconducting loop or $\sigmab$ in our case) coupled to a
dissipative environment. If the particle is initially set up to be
in the lowest state of a single well, in the absence of
dissipation the particle will eventually tunnel between the
minima, with the ground state being a symmetric superposition of
the lowest states of the two wells. In the presence of dissipation
 tunnelling becomes sensitive to the spectral density of the
harmonic oscillator bath at low frequencies. For the case of ohmic
dissipation, Caldeira and Leggett concluded\cite{CL} that
dissipation decreases the tunnelling rate.

Soon after this, Chakravarty\cite{chakravarty} and Bray and
Moore\cite{bray-moore} took the results much further. They pointed
out a deep connection between this problem and the anisotropic
Kondo problem\cite{kondo} of electrons interacting with a
localized spin. They  reached the important conclusion that there
are two phases in the Caldeira-Leggett problem.  At sufficiently
weak dissipation the particle is delocalized between the two
degenerate wells, and this corresponds to the Kondo singlet phase
of the antiferromagnetic Kondo model ($J_{\perp}>0$). At strong
dissipation the particle is localized in one well, corresponding
to the ferromagnetic Kondo model ($J_{\perp}<0$), for which the
spin-flip terms are irrelevant at low energies.

The relation of this work to ours is made more precise if we consider
the effective action\cite{AES} used by Chakravarty\cite{chakravarty} for a
resistively shunted superconducting quantum interference device
(SQUID):
\beqr S_{eff}=&{1\over \hbar}\int\limits_{0}^{\beta\hbar} dt
\left[ {\hbar^2 C\over 8e^2}\bigg({d\t\over
dt}\bigg)^2+u(\t)\right]\nonumber\\ &+2\eta
\int\limits_{0}^{\beta\hbar} {dt\ dt'\over(t-t')^2}
\sin^2\bigg({\t-\t'\over4}\bigg) \label{seff-chakravarty}\eeqr
where $\t$ denotes the flux (in units of $\phi_0/(2\pi)$), $C$ is the
capacitance of the SQUID, $u(\t)$ is the double-well potential
with degenerate minima at $\pm \t_0$, and the dimensionless number
$\eta=\hbar/2\pi e^2 R$ parameterizes the dissipation ($R$ is the
shunt resistance). The phase transition occurs\cite{chakravarty}
(in the limit when the tunnelling amplitude is very small) for
$4\eta \sin^2(\t_0/2)=1$. The effective action for the Kondo
problem\cite{kondo} can also be cast in the above
form.\footnote{The dissipation induced by the bath manifests
itself as a long-range $1/|t-t'|^2$ interaction in Euclidean time.
Exactly the same effective action is obtained by integrating out
the dissipative electrons in the Kondo problem, as was done by
Anderson, Yuval, and Hamann\cite{kondo}. At a physical level, the
particle spends most of the (Euclidean) time in one of the wells,
occasionally tunnelling between them, an event called an
instanton. These instantons correspond to spin-flip events in the
Kondo problem\cite{kondo}.}

In our problem, dissipation at very low frequencies will enter the
effective action through the susceptibility, which is the
quadratic term in $\sigmab$ in the fermionic $Tr Ln$. We have seen
before (Eqn. (\ref{freq-dependence-s0})), in the case of the
isolated dot, that the dynamics of $\sigmab$ become dissipative
for $\omega >> \Delta$ when it can decay into numerous fermionic
many-body states. When connected to the leads, this Landau damping
behaviour persists down to the Fermi energy because the dot states
are now embedded in the continuum and $A_\a^{p,h}$, the spectral
densities of the state labelled by $\a$ above ($p$ for particle)
and below ($h$ for hole) the chemical potential in the exact
basis, evolve from the delta function to a Lorentzian of width
$\Gamma$. \beq A_\a(\ve)=\Theta(\ve)A^{p}_\a(\ve)+
\Theta(-\ve)A^{h}_\a(\ve)={\Gamma\over\Gamma^2+(\ve-\ve_\a)^2}
\eeq Since we are focusing on very low frequencies, the spectral
densities are of order $\Gamma/\Delta^2$. Remembering that
$|M_{\a\b}|^2$ is typically of order $1/g$, and putting in the
overall factor of $g^2$ in the quadratic term, one can easily
check that the imaginary part of the susceptibility which
represents dissipation will therefore be of order $g\Gamma^2 \w$.
Going over to the imaginary time representation, one can show that
up to an unimportant constant we have the following term in the
effective action \beqr
&g{\Gamma^2\over\Delta^2}\int\limits_{-\infty}^{\infty}
{d\w\over2\pi} |\w| \sigmab(\w)\cdot\sigmab(-\w)\cr
&={2g\Gamma^2\langle\sigma\rangle^2\over\pi\Delta^2}\int {dt dt'
\over (t-t')^2} \sin^2\bigg({\chi-\chi'\over2}\bigg)
\label{1overt2}\eeqr In addition  we also have the $\left( {d\chi
\over dt }\right)^2$ term from the higher energy states.

Comparing our full effective action (Eq.
(\ref{eff-action-dynamic}) + Eq. (\ref{1overt2})) to that of
Chakravarty\cite{chakravarty} (Eq. (\ref{seff-chakravarty})), we
see that the angle $\chi$ of $\sigmab$ corresponds to the
``particle'' degree of freedom $\t$, and that
$g\Gamma^2/\Delta^2\approx 1$ corresponds to large dissipation (in
the strong coupling phase $\langle\sigma\rangle$ is of order
unity). Therefore, despite the fact that the two minima are
exactly degenerate, $\sigmab$ will be localized in a single
minimum at large $g$ for $\Gamma>\Delta/\sqrt{g}$. Even when this
condition is satisfied the level width is still much smaller than
the level spacing for large $g$. Under these conditions
time-reversal symmetry will be spontaneously broken for $m$ odd
and zero temperature. As mentioned before, time-reversal symmetry
breaking can be inferred from the statistics of peak
heights\cite{peak-height-th,peak-height-expt}. This
symmetry-breaking also induces a spontaneous persistent current at
zero flux, and will be described in Part II\cite{partII}. Note
however that in order to see this effect we need $T<\Delta
e^{-g}$ since above this $T$, thermal effects will produce the
same effect.

The $\cT$-breaking transition is structurally identical to the
ferromagnetic-antiferromagnetic Kondo transition, but is different
in character from other Kondo-like states proposed in quantum
dots\cite{matveev,DGG}. Any Kondo-like state has to start from a
finitely-degenerate set of states forming a pseudospin, and
``conduction'' electrons with a continuum of energies coupling to
this pseudospin. The two main previous proposals are: (i) The
degeneracy is in the number of particles at the zero-bias
conductance peak, when the dot has the same free energy to have
$N$ or $N+1$ particles, which are the two states of the
pseudospin-$\half$. Electrons hopping from the leads ``flip'' the
pseudospin, and at low enough $T$ form a Kondo resonance at the
Fermi energy\cite{matveev}. (ii) A dot with an odd number of
electrons has a singly-occupied state as the highest occupied
level with spin-$\half$, which then acts as a standard Kondo
impurity\cite{DGG}.  In both these examples, one cannot change the
sign of the Kondo coupling to go from the antiferromagnetic to the
ferromagnetic Kondo model. Thus, while there is an observable
resonance at the Fermi energy below the Kondo temperature $T_K$,
there is no phase transition (see, however, Ref.
\cite{kolomeisky}). The new features of our state are that the
pseudospin is a truly collective variable, which represents the
Fermi surface distortion of all the electrons, and that one can
drive the transition between the analogs of the antiferromagnetic
and ferromagnetic Kondo problem merely by tuning the coupling to
the leads. In contrast to the $g=\infty$ mesoscopic Pomeranchuk
transition which broadened into a crossover for any finite $g$,
this is a true quantum phase transition. The finite system is able
to do this by coupling coherently to the infinite reservoir via
the leads.

\subsection{The Quantum Critical Regime}
\label{critical-fan}

Now we turn to a description of the quantum critical regime, which
can be reached from the strong- or the weak-coupling regimes by
increasing either the probe frequency,gate voltage,  temperature
$T$ or $1/g$. We will focus below on increasing $1/g$.

The most important feature of this regime is that $\sigmab$ has a
gap of order $\Delta$   whereas in the weak-coupling regimes it
has a gap of order $g\Delta $.  This in turn leads to a broad
spectral function for low-energy excited states. This could be
seen in tunnelling at finite bias\cite{sivan-finite-bias}
(nonlinear conductance).

\subsubsection{Connection to Fock-Space Delocalization}

The physics of the quantum critical regime can be connected to
ideas of Fock-Space localization-delocalization
\cite{fock-loc,fock-loc2}, which examine the spectral function of
an excited quasiparticle state in a mesoscopic system as
interactions are turned on. Assuming no correlation between the
matrix elements of the interaction, $V_{\a\b\g\d}$, these studies
concluded   that for excitation energy \beq \ve\ge \ve^*\approx
\Delta \sqrt{g/\ln{g}} \label{estar}\eeq
 the spectral function of the quasiparticle is broadened into a
Lorentzian, characteristic of the infinite system. Below $\ve^*$
the spectral function splits up into a number of sharp peaks. We
briefly recount the  simple arguments behind this  result (modulo
the $\ln{g}$) \cite{fock-loc} as it relates also to our work.
Consider a single particle of energy $\ve$ decaying into a
particle of lower energy and a particle-hole pair by  Fermi's
golden rule. By energy conservation we need only consider the
energy of the particle-hole pair as independent. The matrix
element for the process is $V_{\a\b\g\d}\simeq \Delta/g$, while
the density of states is just the density of particle-hole states
(or two-body states), which is given by the number of ways of
partitioning the total energy $\ve$ among three particles
$\nu_3(\ve)=\ve^2/2\Delta^3$. Note that this is much smaller than
the density of many-body states. The total width induced by this
process is roughly $\Gamma\simeq \ve^2/g^2\Delta$. If this is much
smaller than $\Delta$ one cannot make the continuum approximation,
and the quasiparticle weight will split into a number of sharp
peaks. The correct criterion is to compare the matrix element
between two different three-body states $|u_m|\Delta/g$ to the
typical spacing between such states $1/\nu_3(\ve)$. If the two are
comparable, the quasiparticle weight gets distributed among many
such states. This leads to the criterion $\ve^*\simeq
\Delta\sqrt{g}$. To obtain the extra $\ln{g}$ in Eq. (\ref{estar})
requires a more sophisticated analysis\cite{fock-loc}, in which
Fock-Space is thought of as a lattice, each site of which
corresponds to a single Slater determinant. Hopping on the lattice
is accomplished by the interaction matrix elements $V_{\a\b\g\d}$.
If the system is localized on this lattice, then the quasiparticle
state is comprised of at most a few determinants, and the spectral
function will therefore be a sum of sharp peaks. If the system is
delocalized on this lattice, the quasiparticle state has an
infinite number of Slater determinants, and the spectral function
is broadened into a Lorenztian. To find the threshold energy for
delocalization $\ve^*$ one carries out the analog of the Anderson
locater expansion.

Let us see how this gets modified in our approach wherein  we keep
correlations between different $V_{\a\b\g\d}$ (all of which really
arise from a single Landau parameter). This gives rise to loop
corrections which were neglected earlier\cite{fock-loc,fock-loc2},
which lead at the critical point to $V_{\a\b\g\d}$ which increase
as the energy scale is lowered, as discussed in Section III. In
fact, the Fermi-surface distortion becomes the bosonic dynamical
variable $\sigmab$, which interacts with the fermion to give it a
decay width.

To make this more quantitative, let us consider the retarded,
real-time,  inverse propagator of $\sigmab$ (after self-averaging
the wavefunction sums) in the weak-coupling phase \beq
D^{-1}_{ret}(\omega)={g^2\Delta\over
|u_m|}-{g\Delta^2\over2}\sum\limits_{\a\b}
{N_\b-N_\a\over\ve_\a-\ve_\b-\omega-i\eta} \label{dinv}\eeq where
$\eta$ is a small positive infinitesimal.

\begin{figure}
\narrowtext \epsfxsize=2.4in\epsfysize=2.0in \hskip
0.3in\epsfbox{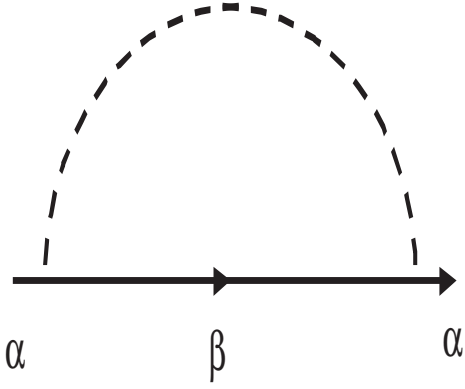} \vskip 0.15in \caption{The lowest-order
self-energy diagram for the fermion due to its coupling to the
collective field $\sigmab$. Straight lines represent fermions
while wavy dotted lines represent $\sigmab$. } \label{self-energy}
\end{figure}

 The fermions couple to $\sigmab$ through the term
$-g\Delta\sigmab\cdot\bM$ (Eq. (\ref{effective})).  The fermion
propagator will get dressed by this coupling, and acquire a
self-energy $\Sigma_{ret,\a}$, the simplest diagram for which is shown
in Fig. (\ref{self-energy}). The decay width produced by this diagram
can be written as
\beqr
\Gamma_{\a}(\w)=&-Im\Sigma_{ret,\a}(\w)=g^2\Delta^2\sum\limits_{\b}
|M_{\a\b}|^2\times\nonumber\\
&\int\limits_{0}^{\w} {d\w'\over\pi}
Im \ D_{ret}(\w-\w') \times  \ Im G_{\beta} (\omega' )
\label{t0-Gamma}\eeqr where $Im$ stands for  the imaginary part of
the function that follows.

In computing the width by the above one loop diagram, we seem to be
following standard diagrammatic perturbation theory. There is,
however, a subtlety that needs to be borne in mind. The $\sigmab$
propagator we use is the not the bare one based on the last term in
the action of Eqn. (\ref{effective}), (where it had no $\omega$
dependence) but an effective one Eqn. (\ref{freq-dependence-s0}), {\em
obtained upon integrating out the fermions}. One may ask what the
fermion is doing in the one-loop graph above if it has already been
integrated out. Is there some double-counting? In Appendix B we will
show that there is no double-counting and that the above procedure for
finding $\Gamma$ is legitimate. Appendix B also explains how to
incorporate symmetry-breaking (a nonzero average for $\sigmab$) into
the calculation.

Resuming our calculation of $\Gamma$, we will assume (and justify
later) that the fermion spectral function is still sharp on the scale
of a few $\Delta$ so that we can make the replacement \beq
Im(G_{ret,\b}(\w))\simeq -\pi\delta(\w-\ve_\b) \label{img0}\eeq inside
the integral of Eq. (\ref{t0-Gamma}). We now consider the spectral
function of the $\sigmab$-propagator.  In Eq. (\ref{dinv}), for
$\Delta\ll\omega\ll E_T$, the $\w$-independent part of the sum over
$\a\b$ produces $g^2/|u_m^*|$, while the dominant $\w$-dependence is
imaginary.  Reading off the imaginary part from Eq. (\ref{dinv}), we
get \beq D_{ret}(\w)\simeq {2/g\pi\over -i\w+\zeta} \label{dret}\eeq
where
\beq \zeta=2rg\Delta/\pi \eeq
 and \beq
r={1\over|u_m|}-{1\over|u_m^*|} \eeq is the distance from the
critical point. Note that this involves replacing the discrete sum
over particle-hole states by a continuum, and will be justified
self-consistently in the following. The imaginary part of this
retarded propagator is \beq Im(D_{ret}(\w))={2\over
g\pi}{\w\over\w^2+\zeta^2} \label{imdret}\eeq This shows that
$\zeta$ acts like a gap or a mass for $\sigmab$, and when $\zeta
\to 0$, $\sigmab$ should become gapless.  Substituting
Eqs.(\ref{img0},\ref{imdret}) into Eq. (\ref{t0-Gamma}) we find
the decay width to be \beq \Gamma(\w)\simeq
{\Delta\over\pi}\ln(1+\w^2/\zeta^2).\label{width-T0}\eeq

In the weak-coupling phase at low energies we have $\w\ll\zeta$.
Expanding the $\ln$ and noting that $\zeta\simeq g/|u_m|$, we
recover the weak-coupling result. At the critical point one must
replace $\zeta$ by $\Delta$. This is the infrared cutoff of the
theory, signifying the energy below which $D_{ret}$ has no
imaginary part. This implies that in the quantum critical regime,
putting the particle on shell ($\w=\ve$), we have \beq
\Gamma(\ve)\simeq {2\Delta\over\pi}\ln{\ve/\Delta}.
\label{width-qc-T0}\eeq

Now consider  for $\Gamma (\ve )$ in the strong coupling phase. We
must expand the theory to quadratic order in $\sigmab$ at the
saddle point.  The fermionic hamiltonian at the saddle point will
take the form \beq
 h_{\a\b} = \delta_{\a\b}\ve_{b}+ g\Delta \sigma_x
 M^{x}_{\a\b}\label{hwithmeansigma}
 \eeq
 where we have chosen the direction of explicit  (sample specific)
 symmetry
 breaking
 to be along the  x-axis. The
 fluctuation about this point have a huge gap of order $g\Delta$
 the $x$ or radial direction
and of order $\Delta$ in the $y$-direction or angular direction.

To proceed, we must first diagonalize $h$ in
Eqn.(\ref{hwithmeansigma}) to get new single particle levels, labeled
$a$, $b$ etc. Their spacing will also be order $\Delta.$ It is the
width of these states that we want to obtain, using the self-energy
diagram of Fig. (\ref{self-energy}).  We need couple only $\sigma_y$ since $\sigma_x$
is very heavy. The coupling matrix will change from $M_x$ to
$U^{\dag}M_xU$, where $U$ relates the $\a, \b$ basis to the $a,b$
basis. We will assume that that the rotated matrices have the same
statistical properties as before rotation, up to factors of order
unity. The exchanged $\sigma_y$ propagator will have a mass of order
$\Delta $ as can be seen from Eq.(\ref{osham}). This results from the
fact that motion in the Mexican Hat (a change of $\sigma_y$) has an
energy cost of only $g\Delta$, as opposed to a cost of $g^2\Delta$ in
the weak-coupling regime. In other words, the Goldstone mode of the
self-averaged theory acquires a small mass due to sample-specific
disorder.  The imaginary part of the $\sigma_y$ propagator will be of
the Landau-damped form and we will obtain a $\Gamma (\omega )$ which
is essentially the same as Eq.(\ref{width-qc-T0}). Thus in entire the
strong-coupling regime, the quasi-particle width is universal (up to
factors of order unity) and of the same form as the quantum critical
region. This is the mesoscopic analog of the Non-Fermi-liquid behavior
found in the bulk Pomeranchuk phase\cite{oganesyan}.

It is very instructive to analyze Eqn.(\ref{width-T0}) from different
points of view and in different limits, since it describes how a very
specific and measurable quantity, the quasiparticle width, is affected
in a real problem where $r$, $g$ and $\Delta$ are at play.

\begin{itemize}
\item{(i)} First, since in a finite
system all many-body states are discrete one should compare the
typical spacing of many-body states with another energy scale (such as
$T$, or the level width due to coupling to the leads, or the finite
energy resolution of the experiment, see Silvestrov\cite{fock-loc2})
to decide whether the width of Eq. (\ref{width-T0}) is apparent as a
continuous spectral function or as a set of sharp peaks with an
envelope given by Eq.  (\ref{width-T0})\footnote{To compute the decay
width at nonzero $T$ we should use the formula \beqr
&Im\Sigma_{ret,\a}(\w)=-g^2\Delta^2\sum_{\a\b}|M_{\a\b}|^2
{1\over\pi}\sum\limits\int d\w' Im(G_{ret,\a}(\w'))\nonumber\\
&\bigg(\Theta(\w'-\w)[N_B(\w'-\w)+N_F(\w')]Im(D_{ret}(\w'-\w))+\nonumber\\
&\Theta(\w-\w')[1+N_B(\w-\w')-N_F(\w')]Im(D_{ret}(\w-\w'))\bigg)
\label{imsigma}\eeqr where $N_F$ is the Fermi function and
$N_B(\w)=(e^{\beta\w}-1)^{-1}$ is the Bose function.} .

\item{(ii)} The crossover between the weak-coupling
and quantum-critical regimes can be seen in Eq. (\ref{width-T0}).
Note that the decay width could have depended on the dimensionless
parameters $r$ and $g$ in an arbitrary way, but in fact does so only
through the energy scale $\zeta=2rg\Delta/\pi$. This behavior is
characteristic of systems near a critical point; the right-hand side
of Eq. (\ref{width-T0}) is a scaling function, and the combination
$\zeta$ is a scaling variable near the quantum phase transition. Since
$g\Delta = E_T$, we see that as one approaches the critical point
$r=0$, the threshold energy for quasiparticle broadening $\ve^*$
decreases as $rE_T$. We emphasize that it never reaches zero; at
$r=0$, it assumes a value $\Delta$, the infrared cutoff. Even if one
is in the weak-coupling regime, one can probe the physics of the
quantum-critical regime by looking at the quasiparticle width at the
energy scale $\ve\simeq\zeta\simeq rg\Delta$. In past
work\cite{fock-loc,fock-loc2}, this behavior was expected at
$\simeq\sqrt{g}\Delta$. Thus when $r$ falls below $1/\sqrt{g}$, the
threshold for such behavior (ill-defined quasiparticle) also drops
linearly with $r$ to a scale as low as $\Delta.$ In units of $E_T$, it
drops from $E_T/\sqrt{g}$ to $E_T/g$.

Since our system has three distinct regimes of behavior, it is
fair ask how one is to navigate around the phase diagram. One
option is to alter the particle density and hence $r_s$. The above
discussion gives us another option: To vary $\omega$ or $\ve$. By
increasing this variable (say by conducting a finite-bias study of
transport\cite{sivan-finite-bias}) we can pass from the
weak-coupling regime to the quantum-critical regime.  By contrast,
in the strong-coupling regime, the width is always as in the
quantum-critical regime. The reason is that the would-be Goldstone
mode $\sigma_y$ has a very small mass (of order $\Delta$) entirely
due to the sample-specific disorder. This is evident from the
hamiltonian of Eqn. (\ref{osham}) describing fluctuations.

\item{(iii)} Let us contrast the underlying physics of the weak-coupling
and quantum-critical regimes.  In the weak coupling limit, $\zeta
=2rg\Delta/\pi$ is very large and we regain \beq \Gamma (\ve ) \simeq
{\ve^2 \over g^2}, \eeq a result we could get directly by
expanding the $\sigmab$-propagator as \beqr D_{ret}(\w)\simeq
&{|u_m|\over g^2}+{2|u_m|^2\over g^3}\times\nonumber\\
&\sum\limits_{\a\b} {N_\b-N_\a\over\ve_\a-\ve_\b-\omega-i\eta} +
\cdots \label{small-um}\eeqr  and carrying through the
$\w'$-integration in Eqn. (\ref{t0-Gamma}). It is clear that
taking higher-order terms in the expansion in powers of $u_m$ lets
the incoming particle decay into final states with higher and
higher numbers of particles/holes. From our previous large-$N$
analysis we know that the radius of convergence of this expansion
is $|u_m^*|$, since there is an instability of the system for
$|u_m|>|u_m^*|$. At the critical point the incoming particle
couples with equal ease to many-body states with
 arbitrarily many particles and holes.

\item{(iv)} The function in  Eq.
(\ref{width-qc-T0}) has no suppression with inverse powers of $g$, in
contrast to the weak-coupling expression $\ve^2/g^2$. This is a result
of the $\sigmab$ gap dropping from order $g\Delta$ to $\Delta$ as
explained in the points made above. On the other hand as a function of
$\ve$ it grows quite slowly. The logarithmic growth of width with
energy justifies the replacement of the fermion spectral function by
the delta function (Eq. (\ref{img0})) since the function it multiplies
in the integral is very broad.

\item{(v)} Recall that in writing down Eq. (\ref{imdret}) we assumed that
the particle-hole states formed a continuum even at low energies.
Armed with the decay rate of Eq. (\ref{width-qc-T0}) we can
justify this self-consistently. The self-consistent $D_{ret}$ is
given by the inverse of Eq. (\ref{dinv}) {\it with the appropriate
self-energies put in for $\ve_\a$ and $\ve_\b$}. Thus, each sharp
pole in $D^{-1}_{ret}$, formerly at $\ve_\a-\ve_\b$ will now be
smeared out over an energy range of order $\Gamma_\a+\Gamma_\b$.
If this width is greater than the typical spacing of two-body
states (which is $2\Delta^3/\ve^2$) the continuum approximation is
valid. It is easy to verify that this holds for energies as low as
a few $\Delta$ (modulo the caveat in point (i)).

\item{(vi)} In the quantum-critical and strong-coupling regimes
we have argued that single-particle states are strongly connected to
many-body states at low energies. One signature of this would be the
statistics of the many-body states. When a single Slater determinant
forms a good description of the ground state, one expects the
single-particle levels to obey Wigner-Dyson statistics, while the
low-energy many-body states are nearly uncorrelated and obey Poisson
statistics. However, in the quantum-critical and strong-coupling
regimes we expect the many-body states at low energies (a few $\Delta$
but independent of $g$) to be strongly coupled and therefore obey
Wigner-Dyson statistics. In fact, an exact diagonalization
study\cite{reentrant} is consistent with this picture. Berkovits and
Avishai\cite{reentrant} find that at small coupling strength the
many-body states display Poisson statistics, while at intermediate
coupling strength they obey Wigner-Dyson statistics. They also find a
reentrance of Poisson statistics at very high couplings, possibly a
signature of a Wigner glass regime\cite{wigner-like-th,shklovskii}.
\end{itemize}

The decay width of the quasiparticles at nonzero energy above the
Fermi energy should be visible in nonlinear conductance
experiments\cite{sivan-finite-bias} (conductance at finite bias).

Finite-$T$ effects in the weak-coupling regime
are worth a little more elaboration. Assume that one is in the
weak-coupling regime, whose physics is dominated by the Universal
Hamiltonian\cite{H_U}. However, there might be observable effects
associated with quantum and/or thermal fluctuations of $\sigmab$ on
the peak-height distribution. The experiment\cite{peak-height-expt2}
shows a narrower distribution than that expected for noninteracting
electrons. Many ideas have been offered to account for this
discrepancy, including dephasing\cite{pk-ht-dephasing}, the effect of
completely random two-body interactions\cite{pk-ht-random-int},
spin-orbit interactions\cite{pk-ht-spin-orbit}, and finite temperature
and exchange\cite{pk-ht-T-exchange,yoram}. While all these may be
contributing factors, the analysis of Usaj and
Baranger\cite{pk-ht-T-exchange} and Alhassid {\it et al}\cite{yoram}
including the effects of exchange in the Universal
Hamiltonian\cite{H_U} sense, and the effect of nonzero $T$, seems to
be the most comprehensive. However, despite the fact that a larger
value of the exchange constant is used\cite{pk-ht-T-exchange} than
predicted by RPA (which makes the distributions narrower) the
experimental distribution is even narrower than the prediction, and
the discrepancy increases with increasing $T$.

In our approach, as $T$ increases, higher energy states of both
the fermions and $\sigmab$ are populated, which will modify the
imaginary part of the fermion Green's function (Eq.
(\ref{imsigma})). This in turn can serve as the input to the
formula derived by Meir and Wingreen\cite{meir-wingreen} for the
current through a fully interacting region, which in the simplest
case of equal couplings to the right and left leads reduces to
\beq I=-{2e\over h} \int d\ve
(N_{F,L}(\ve)-N_{F,R}(\ve))Im(Tr[\bGamma \bG]) \eeq where $N_F$
are the Fermi functions of the two leads,
$\bGamma=\bGamma_L\bGamma_R/(\bGamma_L+\bGamma_R)$ is the level
width matrix and $\bG$ is the matrix single-particle Green's
function (including all many-body contributions).

It is qualitatively clear that as one approaches the quantum
critical regime by increasing $T$ or decreasing $\zeta$, the
single-particle fermionic states get more and more strongly
coupled to many-body states. This will lead to incoherent
averaging of the conductance and thus a narrowing of the peak
height distribution beyond that implied by the Universal
Hamiltonian treatments\cite{pk-ht-T-exchange,yoram}. A
quantitative calculation is beyond the scope of this paper.

\subsection{The Order Parameter and its Fluctuations}

We are now in a position to examine the size of the order parameter
and its fluctuations in the three different regimes, and to delineate
the crossovers between them. Once again, even and odd $m$ turn out to
be significantly different, and we will treat them in turn.

In the case of even $m$ a term linear in $\sigmab$ is allowed, and the
effective potential is generically of the form
\beq
V_{eff}(\sigmab)=-g\sigma\Delta\cos(\chi-\chi_0) +\half g^2\Delta r\sigmab^2
+{\lambda\over4} g^2\Delta (\sigmab^2)^2+ \cdots
\eeq
where the first term arises from the realization-specific disorder and
the other terms are self-averaging. The $\cos(\chi-\chi_0)$ is
intended to show the explicit breaking of the rotational invariance in
$\sigmab$-space by the sample-specific disorder. By keeping only the
terms kept above, and minimizing in the magnitude and angle of
$\sigmab$, it is easy to verify that near the critical point ($r\ll
1$) the average value of the order parameter is the solution to a
cubic and has the form
\beq
\langle\sigma\rangle\simeq g^{-1/3} f(rg^{2/3})
\label{order-evenm}\eeq
where $f(x)$ is a scaling function which becomes a constant as $x\to0$
and behaves as $\sqrt{|x|}$ for large negative $x$. Thus the order
paramter behaves as $g^{-1/3}$ in the quantum-critical regime and
crosses over to a $\sqrt{|r|}$ behavior deep in the strong-coupling
regime. From Eq. (\ref{order-evenm}) it is clear that the crossover
between the weak-coupling and quantum-critical regimes happens for
$|r|g^{2/3}\approx 1$ as far as the order parameter is concerned. This
applies equally to the quantum-critical to strong-coupling
crossover. Note that this is a different crossover than that found in
quasiparticle broadening.

For odd $m$, only even functions of $\sigmab$ are allowed in
$V_{eff}$, even among the realization-specific terms. The effective
potential  is generically
\beq
V_{eff}(\sigma)=-g\Delta \sigma^2 \cos{2(\chi-\chi_0)}+\half g^2\Delta r\sigmab^2
+{\lambda\over4} g^2\Delta (\sigmab^2)^2+ \cdots
\eeq
For simplicity, we will assume that either a small nonzero external
flux selects one of the two degenerate minima in the following.
Clearly, the crossover to the quantum-critical regime happens when the
self-averaging quadratic term is comparable to the
realization-specific term, leading to the criterion $rg\approx 1$. In
the quantum-critical regime the order parameter size is of order
$1/\sqrt{g}$, while deep in the strong-coupling regime it is behaves
as $\sqrt{|r|}$.

Let us now look at the fluctuations of the order parameter. In
previous work\cite{qd-us2} two of us had wrongly stated that
quasiparticle broadening was connected to the delocalization of
$\sigmab$. As has been discussed in the previous subsection at some
length, quasiparticle broadening in the quantum-critical and
strong-coupling regimes is a direct result of the small gap in the
$\sigmab$ propagator (of order $\Delta$). However, the fluctuations of
$\sigmab$ remain bounded and nearly the same in all three regimes. To
see this, let us calculate
\beq
\langle\sigmab^2(t)\rangle=\int\limits_{0}^{g\Delta}{d\w\over2\pi} D(\w)
\eeq
where $D$ is the Euclidean propagator of $\sigmab$.  Here we have
cut off the frequency integration at $E_T=g\Delta$, because retaining
higher frequencies would be inconsistent with our starting point, in
which all states of energy higher than $E_T$ have been integrated
out. From Eq. (\ref{dret}) we obtain
\beq
D(\w)={2/g\pi\over|\w|+\zeta}
\eeq
and thus
\beq
\langle\sigmab^2(t)\rangle\simeq {2\over g\pi} \log\bigg[{g\Delta\over\zeta}\bigg]
\eeq
In the weak-coupling regime $\zeta\simeq rg$, and the size of the
$\sigmab$ fluctuations is $1/\sqrt{g}$. This increases a little in the
quantum-critical and strong-coupling regimes (where
$\zeta\simeq\Delta$) to $\sqrt{\log{g}/g}$. To reiterate, Fock-space
delocalization has to do with the {\it gaplessness} of $\sigmab$, and
has no relation to its fluctuations.

\subsection{Summary of Signatures and Connection to Experiments}

Since this section has been rather long and involved, it is
appropriate to collect in this subsection the signatures of the
existence of the collective variable in the various regimes. We remind
the reader that we have restricted attention to charge-channel
instabilities for spinful electrons, in which the spin index is a
passive spectator. This restriction is not unreasonable in the light
of the values of the Landau parameters\cite{kwon-ceperley} for $r_s=5$
(Eq. (\ref{fli-parameters})). We emphasize that only the signatures
pertaining to Coulomb Blockade are discussed in detail in this
paper. Other equally striking signatures in the persistent current
will be explored exhaustively in Part II\cite{partII}.

\begin{itemize}
\item{} Strong-coupling regime for all $m$: $\sigmab$ is localized
in a single minimum for $m$ even, or a pair of minima related by
time-reversal for $m$ odd, with the size of $|\sigmab|$ being of order
unity. The addition spectrum and therefore the conductance
peak-spacing distribution are broader and more symmetric than the GOE
level spacing distribution. This results from rare large shifts of
$\sigmab$ as particles are added, which also leads to a correlation
between small peak-spacing and exponentially small peak height. The
$m$ even case has a diamagnetic persistent current\cite{short} in
response to an external flux, while the $m$ odd case has a
paramagnetic persistent current. Low-energy quasiparticles are
broadened throughout the strong-coupling regime, as will be elaborated
below.

\item{} Strong-coupling regime for $m$ odd, isolated dot: At $T=0$,
the system is in a symmetric superposition of two exactly degenerate
minima of the effective potential. However, the crossover flux is
exponentially small in $g$, of the order of $\phi_{cross}\approx\phi_0
e^{-g}/g$. The breaking of time-reversal by this tiny flux can be
experimentally monitored in the peak-height
distribution\cite{peak-height-th,peak-height-expt}, which corresponds
to the GUE for $\phi>\phi_{cross}$. In the flux regime
$\phi_0/\sqrt{g}\gg\phi\gg\phi_{cross}$ the peak position varies
linearly with $\phi$, in contrast to the noninteracting case in which
it would vary quadratically.

\item{} Strong-coupling regime for $m$ odd,  dot  coupled to leads:
The dynamics of $\sigmab$ becomes dissipative at arbitrarily low
energies, and our problem can be mapped on to the Caldeira-Leggett
problem\cite{CL}. From its solution\cite{chakravarty,bray-moore}, we
conclude that at sufficiently high dissipation, corresponding to level
widths (due to the dot-lead coupling) of $\Gamma\simeq
\Delta/\sqrt{g}$, the system can be driven to spontaneously break
time-reversal symmetry. This transition can once again be
experimentally accessed via the peak-height
distribution\cite{peak-height-th,peak-height-expt}, and is isomorphic
to a ferromagnetic to antiferromagnetic Kondo transition.

\item{} Quasiparticle width: Throughout the quantum-critical and
strong-coupling regimes $\sigmab$ becomes ``gapless'' in the scaling
limit ($g\to\infty,\ \ \Delta\to0$ with $E_T$ fixed). A
single-particle description is no longer valid even at low energies
(of order a few $\Delta$), and low-energy quasiparticles are
``Fock-space delocalized''. This prediction is in contrast to previous
work\cite{fock-loc,fock-loc2} which suggested a broad quasiparticle
only at energies of order $\simeq \Delta\sqrt{g}$. Physically this
corresponds to strong mixing between the low-energy single-particle
excitations and collective overdamped excitations of $\sigmab$. In the
quantum-critical regime both vector components of $\sigmab$ are
gapless, while in the strong-coupling regime, the fluctuations of
$\sigmab$ in the nearly degenerate Mexican Hat potential (the would-be
Goldstone mode) are nearly gapless.

\item{} Weak-coupling regime: Despite the fact that it is
dominated by Universal Hamiltonian physics, quantum and/or thermal
fluctuations of $\sigmab$ might have a measurable effect on the width
of the peak-height distribution. Such effects are predicted to get
stronger as one goes towards stronger coupling (by increasing $r_s$).
The quantum-critical regime can be accessed by measurements at finite
bias, nonzero  frequency, or nonzero $T$.
\end{itemize}

Let us now briefly analyze the relevant experiments. The Sivan
{\it et al}\cite{small-rs-expt} and Patel {\it et
al}\cite{small-rs-expt} experiments are done on gated GaAs 2DEG
samples, for which $r_s\approx 1-1.2$ in the bulk. Using the area
of the sample and the fact that these dots are in the ballistic
limit we can find both $\Delta$ and $E_T$, and thence find
$g\approx 7-14$. Sivan {\it et al}\cite{small-rs-expt} find that
the CB peak spacing is about 5 times broader than that predicted
by the Universal Hamiltonian, which describes the weak coupling
region of our phase diagram. However, Patel {\it et
al}\cite{small-rs-expt} find it to be in accord with this
prediction, after accounting for ``experimental noise'' which is
determined by measuring the magnetic field asymmetry of the CB
spacings (which indicates motion of the dopant atoms, or some
other scrambling of the single-particle potential).  Thus, the
Patel {\it et al} data seem to lie in the weak-coupling region.
The Sivan {\it et al} data also correspond to the weak-coupling
region. Presumably, if the subtraction of experimental noise were
to be carried out this data would also match the Universal
Hamiltonian predictions.

The experiments by Simmel {\it et al} and Abusch-Magder {\it et
al}\cite{large-rs-expt} are performed on Si quantum dots with
$r_s\approx2.2$, and $g\approx 18$. (For $r_s\ge2$ local charge
density correlations develop in the dot\cite{wigner-like-th},
similar to the classical limit\cite{shklovskii}. While a Fermi
surface distortion is not a charge density wave, it enhances the
susceptibility for one, and could thus be a precursor). Two
signatures of the critical fan are found in this
experiment\cite{large-rs-expt}. The CB peak-spacing distribution
is found to be 7-8 times wider than expected from the Universal
Hamiltonian (assuming spin degeneracy). Also, the width of the CB
peaks does not vanish\cite{large-rs-expt} as $T\to0$. This is just
what is expected for a system was located in the critical fan: The
ground and low-lying excited states are ``Fock-space
delocalized'', and single-particle states are broad even at low
energy. However, it is not clear whether the width of the peaks is
intrinsic or dominated by charge rearrangement
noise\cite{kastner-private}.

Finally, as has been discussed in the previous subsection,
measurements of the width of the peak-height
distribution\cite{peak-height-expt2} and its evolution with $T$
are not completely understood from the Universal Hamiltonian
standpoint\cite{pk-ht-T-exchange,yoram}. Our theory predicts a further
narrowing of the peak-height distribution.

In conclusion, current experiments seem to be in the weak-coupling
regime, with possible excursions into the quantum-critical crossover
at nonzero temperatures. From the values of the Landau parameters for
the clean 2DEG\cite{kwon-ceperley} one can guess that systems with
$r_s$ substantially larger than 5 will be needed to see the
strong-coupling regime, and that charge-channel instabilities are
likely to occur first, although occasionally even systems with $r_s=5$
may happen to be in this regime (see the end of Appendix A). In
constructing more strongly interacting sample, one must bear in mind
that since $g\simeq k_F L$, reducing the density at constant dot size
$L$ will reduce $g$. What one would ideally like to do is to keep $g$
fixed, which means preparing larger dots while reducing electron
density.

\section{Conclusions and Open Questions}

Starting with the problem of interacting two-dimensional electrons in
a ballistic mesoscopic structure with chaotic boundary scattering, we
found that Landau Fermi-liquid interactions (parameterized by
dimensionless Landau parameters $u_m$ in the $m^{th}$ angular momentum
channel) were the most natural starting point at the Thouless scale
$E_T$. The only condition for this to hold is that $E_F/E_T$ be
large. For a ballistic/chaotic quantum dot this ratio is of the same
order as the dimensionless conductance $g=E_T/\Delta$, where $\Delta$
is the mean level spacing. Thus the largeness of $g$ is a sufficient
condition for Fermi liquid interactions to be applicable in the
Thouless band. This is a new regime in the broad problem of the
interplay of interactions and disorder. The special features of this
regime which make it controllable are the applicability of Random
Matrix Theory, and the presence of the small parameter $1/g$, which
conspire to make a large-$N$ ($N=g$) approximation feasible. The
solution is nonperturbative in both interaction strength and disorder:
The disorder is the strongest it can possibly be, scattering momentum
states chaotically into each other, while the interactions have a
dimensionless strength of order unity. The secret to the success of
the nonperturbative solution is the additional small parameter $1/g$
which controls the size of fluctuations in the large-$N$
approximation.  Note that in contrast to many applications in which
the large-$N$ approach is used, in this case it is clearly justified
by the largeness of $g$ (5-20 in experimental samples).

Restricting our attention to charge-channel instabilities, we found a
disordered version of the Pomeranchuk shape transition in every Landau
Fermi-liquid channel, confirming the one-loop RG result found earlier
by two of us\cite{qd-us1}, and putting it on a rigorous footing as
being exact in the large-$g$ limit\cite{qd-us2}. The corresponding
Landau parameter had to be $u_m\le u_m^*=-1/2\ln{2}=-0.7213$, as
compared to the bulk instability at $u_m\le-1$. Thus there is a window
in which the clean bulk system is stable, but the mesoscopic
ballistic/chaotic system is not, which indicates that both
interactions and disorder are crucial to the existence of this
regime. More specifically, we saw that interactions produce a Mexican
Hat potential, and the realization-specific disorder allowed
symmetry-breaking even in the presence of finite-$g$ quantum
fluctuations. This can be understood as explicit mass generation for
the would-be Goldstone mode by sample-specific disorder.

The transition at $g=\infty$ is replaced by a crossover for any finite
$g$. In accord with generally known features of quantum phase
transitions\cite{critical-fan} the coupling constant-$1/g$ plane is
divided into a weak-coupling regime, a strong-coupling regime, and a
fan-shaped quantum critical regime. However, our theory displays
additional nontrivial features not found in clean models, such as the
persistence of symmetry-breaking even in the zero-dimensional limit,
due to sample-specific disorder. In particular, in our theory the
shape of the crossover regime depends on the property being
considered: For example, the quasiparticle width at low energies has
only a single crossover between the weak-coupling and quantum-critical
regimes and remains broad throughout the strong-coupling regime. On
the other hand, the order parameter displays an additional crossover
from the quantum-critical to the strong-coupling regime, with the
shape of the crossover depending on whether the angular momentum
channel $m$ in which the instability occurs is even or odd. It must be
reiterated that these properties can be explicitly calculated in a
controllable large-$N$ approximation in all the regimes thanks to the
small parameter $1/g$.

Note that we make no attempt to keep all $1/g$ contributions in the
effective action $S_{eff}$, instead keeping only the lowest nontrivial
terms of each type. As an example, we do not keep $1/g$ corrections to
the self-averaging part of the effective action, but we do keep the
$1/g$ terms which break the rotational invariance in $\sigmab$-space
of the self-averaged part, since they have a profound effect on the
physics of the strong-coupling regime. Similarly, we keep the
``kinetic'' terms in $S_{eff}$ which are down by $1/g$ compared to the
self-averaging static part of $S_{eff}$, since these are the lowest
order nonzero terms.

The strong-coupling regime is where the most striking properties of
the collective Fermi surface distortion field $\sigmab$ are
found. Here a bifurcation occurs with some properties being common to
both even and odd $m$, while others depend on $m$ modulo 2. The
features shared by all $m$ in the strong-coupling regime include: (i)
A mean value of the order parameter of order unity. (ii) Rare large
shifts of the order parameter upon adding a particle, leading to a
broadened peak-spacing distribution. (iii) The same phenomenon as (ii)
leading to a correlation between small peak-spacing and exponentially
small ($\simeq\exp{-g\ln{g}}$) conductance due to the
near-orthogonality\cite{parametric,orthogonality} of the states at
different $\langle\sigmab\rangle$. (iv) Quasiparticles become broad at
energies of a few $\Delta$ throughout the strong-coupling regime. A
special property of the $m$ even strong-coupling regime is a
diamagnetic persistent current\cite{short}, to be investigated in
greater detail in Part II\cite{partII} of this paper.

The strong-coupling regime of a system with an instability in an odd
$m$ channel shows even more striking effects, which depend sensitively
on how strongly the dot is coupled to the leads. The physics can be
mapped on to the Caldeira-Leggett model\cite{CL}, and is driven by the
combination of an exact two-fold degeneracy in the effective potential
(a consequence of the time-reversal invariance of the underlying
Hamiltonian) and the low-energy dissipation induced by coupling to the
leads. At weak dot-lead coupling, the system is ``almost'' in a
time-reversal- ($\cT$-) broken state. The crossover flux need to tip
the system into a completely $\cT$-broken regime is exponentially
small ($\phi_{cross}\simeq\phi_0 e^{-g}/g$), as contrasted with a
crossover flux of $\phi_0/\sqrt{g}$ needed for a noninteracting
system. When the dot-lead coupling increases to a value such that the
broadening of single-particle states due to the leads reaches
$\Gamma\simeq \Delta/\sqrt{g}$, a spontaneous breaking of
time-reversal takes place\cite{chakravarty,bray-moore}. This is a true
quantum phase transition in the universality class of the
antiferromagnetic to ferromagnetic Kondo model\cite{kondo}, and can be
acheived by a finite system by coupling coherently to an infinite
reservoir. In the $\cT$-broken phase the system also displays a
spontaneous paramagnetic persistent current\cite{short} of order
$E_T/\phi_0$, to be explored in greater detail in Part
II\cite{partII}.

Coming to the quantum-critical regime, the main feature is the
``gaplessness'' of the collective variable $\sigmab$, by which we
mean that its propagator has the Landau-damped form $-1/i\w$ for
$\Delta\ll\w\ll E_T$. In this regime low-energy single-particle
excitations are strongly coupled to many-body excitations with
arbitrary numbers of particle-hole pairs via the collective
variable $\sigmab$. This has many consequences: (i) Recall that
the threshold for a continuous and broad spectral function for a
particle was found to be $\ve^*\simeq \Delta\sqrt{g/\ln{g}}$ in
earlier work\cite{fock-loc,fock-loc2} which ignored correlations
between interaction matrix elements, and therefore missed the
quantum phase transitions found here. In our approach we find that
this threshold $\ve^*$ decreases to a few $\Delta$ {\it
independent of $g$} in the quantum-critical regime. This could
potentially be seen in nonlinear conductance (conductance at
finite bias) experiments\cite{sivan-finite-bias}. (ii) Since they
are strongly mixed by interaction effects, the statistics of the
low-energy many-body states should change in the quantum-critical
regime from Poisson towards Wigner-Dyson, as has been seen in
recent numerics\cite{reentrant} (iii) Finally, and perhaps most
relevant for current experiments, one should be able to access the
quantum-critical regime from the weak-coupling regime by
finite-bias, finite-frequency, or finite-temperature measurements.
In particular, there should be a narrowing of the peak-height
distribution beyond Universal Hamitonian
effects\cite{pk-ht-T-exchange,yoram} at nonzero temperature.

As has been mentioned above, the broadening of low-energy
quasiparticles persists in the strong-coupling regime due to the
near-gaplessness of the would-be Goldstone mode of the Mexican
Hat. This seems to be the mesoscopic analog of the non-Fermi-Liquid
behavior found throughout the bulk Pomeranchuk phase in clean
systems\cite{oganesyan}.

Finally, as regards the weak-coupling regime, the collective variable
$\sigmab$ has a huge gap here of order $rg$
($r={1\over|u_m|}-{1\over|u_m^*|}$ is the distance from the critical
point). Thus the weak-coupling regime is controlled by the Universal
Hamiltonian\cite{H_U} of Eq. (\ref{hu}), which is the low-energy fixed
point in this regime.

The picture that emerges is that just as Landau's Fermi liquid theory
is unstable to the bulk Pomeranchuk instability in any Fermi liquid
channel, the Universal Hamiltonian is also unstable to the mesoscopic
Pomeranchuk instability in any channel, with the instability occuring
more readily in mesoscopic ballistic/chaotic systems than in the
bulk. The mesoscopic Pomeranchuk regimes represent physics
nonperturbatively different from Universal Hamiltonian physics. These
strong-coupling regimes, and the crossover quantum-critical regimes,
can be accessed by changing the electron density, or temperature, or
the probe frequency. The consilience of Random Matrix Theory and
large-$g$ enables us to carry out controlled calculations and make
predictions in all three regimes. Our analysis provides a new
framework for analysing future experiments on strongly interacting
mesoscopic systems.

Let us now turn to open questions and possible future directions.
Apart from a more precise treatment of finite-$T$
effects\cite{usaj-finite-T,pk-ht-T-exchange,yoram} in our theory, the
most pressing open question is the treatment of spin-channel
instabilities, which necessarily involves the inclusion of the
Universal Hamiltonian exchange coupling\cite{H_U} $J$. In other words,
what is the interplay between the mesoscopic Stoner
transition\cite{H_U} and the mesoscopic Pomeranchuk transition in a
spin channel? One can imagine a cooperative phenomenon whereby the
exchange coupling favors symmetry-breaking in a spin Fermi-liquid
channel. This might reduce the magnitude of the critical coupling
$Z_m^*$ considerably, and hence make the transition observable at
smaller $r_s$. Exotic types of ordering intertwining spin and momentum
space indices might also occur\cite{varma}. 

A second effect which deserves investigation is the following: We have
assumed that no renormalization of the Landau parameters happens at
energy scales higher than $E_T$, whereas they do renormalize below
$E_T$ in the manner described in Section III.  Clearly, there must be
some renormalization of the Landau parameters above $E_T$, and since
the sign of the flow cannot abruptly change, this effect also tends to
reduce the critical coupling. To rephrase the question, are the Landau
parameters of the chaotic/ballistic cavity the same as those in the
bulk 2DEG of the same electron density? Perhaps the best way of
resolving this issue is to carry out the analog of the bulk quantum
Monte Carlo calculations\cite{kwon-ceperley} directly in a
chaotic/ballistic cavity\cite{HUB}.

A somewhat related question concerns slightly diffusive quantum
dots, where the bulk mean free path $l$ is smaller than (but not
negligible compared to) the size of the dot $L$. In this case
there is an energy regime between $\hbar v_F/L$ and the Thouless
energy $E_T=\hbar D/L^2$. In this intermediate energy regime there
will be a crossover between bulk diffusive dynamics and Random
Matrix dynamics. In the extreme diffusive limit one must recover
the Finkelshtein scaling\cite{int+disorder,belitz}.

One can also ask what happens as one increases the interaction
strength (or $r_s$) to very large values. Is there physics beyond the
strong-coupling regimes we have found? Certainly for the Coulomb
interaction we know that one must eventually cross over into a Wigner
Glass (or Coulomb Glass) regime\cite{wigner-like-th,shklovskii}. The
flattening of parts of the Fermi surface in the mesoscopic Pomeranchuk
regimes increases the susceptibility for charge density wave
instabilities. This indicates that the crossover to the extremely
strong-coupling Wigner Glass regime might happen more readily in
mesoscopic ballistic/chaotic samples than in bulk disordered
samples. An investigation of this crossover would complete the picture
of strong-coupling physics in mesoscopic ballistic/chaotic samples.

At a more philosophical level, the identification of a new,
controllable, regime of disorder and interactions opens up many
possibilities. As an example, one could apply our RG techniques
with RMT correlations to the Kondo problem\cite{kondo}, and
investigate the effect of Landau Fermi-liquid interactions between
conduction electrons residing in a ballistic/chaotic
cavity\cite{kondo-box} on the Kondo effect. Another example to
which our techniques might be applied is the effect of
Fermi-liquid interactions on granular, disordered, gapless
superconductors exhibiting a novel metal-insulator
transition\cite{sfbn}, since the eigenvalue and eigenvector
statistics of their grains are expected to be governed by one of
four newly discovered RMT universality classes\cite{zirnbauer}.

To conclude, ballistic/chaotic quantum dots seem to provide us with a
unique theoretical and experimental playground with exquisite control
on both fronts. On the theoretical front, RMT and the $1/g$ expansion
allow us to tackle disorder and interactions simultaneously and
nonperturbatively.  On the experimental front dots give us the best of
both worlds: A phase transition at $g=\infty$ which we can "see" even
at finite $g$ and access to all the control parameters a mesoscopic
system allows (varying $r_s$, external flux, bias voltage, and
coupling to leads).

We are grateful to the NSF for grants DMR-0071611 (GM), and DMR-
0103639 (RS), DMR-98-04983 (DH and HM), the Aspen Center for
Physics for its hospitality, and Yoram Alhassid, Harold Baranger,
Sudip Chakravarty, Claudio Chamon, Yuval Gefen, Bert Halperin,
Yong-Baek Kim, Eduardo Mucciolo, Chetan Nayak, Yuval Oreg, Zvi
Ovadyahu, Boris Shklovskii, Doug Stone, Denis Ullmo, and Chandra
Varma for illuminating conversations.

\section{Appendix A}

In this appendix we will show that the RG equation and the
quadratic part of the effective action are self-averaging. Let us
first focus on the RG, and look at the part of Eq.
(\ref{diagram-leading}) which is to be summed over \beqr
R=&\sum\limits_{\nu=-\Lambda}^{0}
{1\over\Lambda+|\ve_\nu|}\sum\limits_{\bp\bp'}
u(\t_\bk-\t_\bp)u(\t_{\bp'}-\t_{\bk'})\nonumber\\
&\times\phi^*_\mu(\bp)\phi^*_\nu(\bp')\phi_\nu(\bp)\phi_\mu(\bp')
\label{to-be-ave}\eeqr As has been stated before, in RMT the
wavefunction averages do not depend on the energy separations of
the states, so the wavefunction average can be carried out
separately. This average in the GOE for $\mu\ne\nu$ and generic
momentum labels is Eq. (\ref{crucial1}), reproduced below for
convenience

\beqr
<\phi^*_\mu(\bp_1)\phi^*_\nu(\bp_2)\phi_\nu(\bp_3)\phi_\mu(\bp_4)>=\nonumber\\
{\d_{14}\d_{23}\over g^2}-{\d_{13}\d_{24}+\d_{1,-2}\d_{3,-4}\over
g^3} \label{crucial1-appA}\eeqr

The $1/g^2$ term is the ``naive Wick contraction'' of
leading-order RMT, but the $1/g^3$ term is necessary to maintain
orthogonality between $\mu$ and $\nu$. The final term would be
missing in the GUE. Substituting the correct momentum labels for
the particle-hole diagram we see that the wavefunction average is

\beq {\d_{\bp\bp'}\over g^2}-{1+\d_{\bp,-\bp'}\over g^3}
\label{crucial2-appA}\eeq Let us first carry out the ensemble
average of $R$. Using Eq. (\ref{crucial2-appA}) in Eq.
(\ref{to-be-ave}) we obtain a convolution of the two Fermi liquid
functions \beq
\sum\limits_{\bp}u(\t_\bk-\t_\bp)u(\t_\bp-\t_{\bk'})=
g\big(u_0^2+\half\sum\limits_{m=1}^{\infty}
u_m^2\cos{m(\t-\t')}\big) \label{convolve-appA}\eeq where we have
reverted to the notation $\t=\t_\bk,\ \t'=\t_{\bk'}$. Note that we
have used \beq \sum\limits_{\bp} =g\int {d\t_\bp\over 2\pi} \eeq
which is valid for large $g$. In the second term of Eq.
(\ref{crucial2-appA}), the $\d_{\bp,-\bp'}$ turns out to be
subleading, while the other allows independent sums over $\bp,\
\bp'$. This means that only $u_0$ contributes due to this term,
which produces

\beq
\sum\limits_{\bp\bp'}u(\t_\bk-\t_\bp)u(\t_{\bp'}-\t_{\bk'})=g^2
u_0^2 \label{convolve-u0-appA}\eeq

Next we need to do the energy sum, which we replace by an integral
\beq
\sum\limits_{\ve_\nu=-\Lambda}^{0}{1\over\Lambda+|\ve_\nu|}\approx
\int\limits_{0}^{\Lambda}{d\ve\over\Delta}{1\over
\Lambda+\ve}={\ln{2}\over\Delta} \eeq

Putting all the pieces together, we obtain the ensemble average of
$R$ to be \beq \ll R\gg = {\ln{2}\over 2g\Delta}
\sum\limits_{m=1}^{\infty} u_m^2 \cos m(\t-\t') \eeq

Now consider the fluctuation of the quantity $R$ of Eq.
(\ref{to-be-ave}), that is, $R-\ll R\gg$. Each term in $R$ is of
order $1/g^3$, where one factor of $1/g$ comes from the energy
denominator, while each wavefunction contributes $1/\sqrt{g}$.
There are $g^3$ such terms coming from the three summations. Since
the average has been taken out, each contribution is completely
random and comes with a random sign. From a one-dimensional random
walk argument, with each step being of size $1/g^3$, and $g^3$
steps, the net answer must be of order \beq \bigg(R-\ll
R\gg\bigg)\simeq {1\over g^{3/2}}<<\bigg(\ll R\gg\bigg) \eeq

Thus we see that the average of $R$ indeed dominates its
fluctuations over the ensemble, and $R$ is therefore
self-averaging.

The situation is simpler for the quadratic term of the effective
action, Eq. (\ref{quadratic}) (call it $S_2$). \beqr S_2&=&g^2
\sum\limits_{\a\b}{N_F(\b)-N_F(\a)\over\varepsilon_\a-\varepsilon_\b}
\sum\limits_{\bk,\bk'}\phi_{\a}^*(\bk)\phi_\b(\bk)\phi_{\b}^*(\bk')\phi_\a(\bk')
\nonumber\\ &   & \!\!\!\!\!
(\sigma_1\cos{m\theta}+\sigma_2\sin{m\theta})(\sigma_1\cos{m\theta'}+\sigma_2\sin{m\theta'})
\eeqr

In effect, $S_2$ is the same as $R$, except for the fact that (i)
only a single Fermi liquid channel contributes, and (ii) there is
an extra sum over $\mu$, which was not present in $R$. The
self-averaging of $S_2$ follows the same lines as above. Because
of the extra sum (and the factor of $g^2$ in front) the ensemble
average $\ll S_2\gg$ is of order $g^2$. Considering the same
random-walk argument for $S_2-\ll S_2\gg$ we have four sums ($g^4$
steps in the random walk) with each step being of order $1/g^3$,
and an additional factor of $g^2$ in front, leading to \beq
\bigg(S_2-\ll S_2\gg\bigg)\simeq g^2 {\sqrt{g^4}\over g^3}\simeq
g<< \bigg(\ll S_2\gg\bigg) \label{self-ave-s2}\eeq Thus the
fluctuations of $S_2$ are $1/g$ down from its self-average.

There is one subtlety with the effective action which is not
present in the RG equation. The denominators of the RG equation
can never be smaller than $\Lambda$, while those of $S_2$ go all
the way to the smallest gap near the Fermi energy. In the GOE the
ensemble average $\ll S_2^2\gg$ formally diverges due to the
linear behavior of the level spacing distribution for small
spacings. This implies that while the distribution for $S_2$ is
sharply peaked around its average, it also has a long tail, which
implies that the critical coupling $u_m^*$ has a long tail as
well. Calling the change in the critical coupling $\d u^*$, and
taking only the smallest level spacing into account, we find the
tail to be \beq p_{tail}(\d u^*)\approx 1/g^2(\d u^*)^3 \eeq This
has the consequence that even systems nominally in the
weak-coupling regime may occasionally be in the strong-coupling
regime, with a probability that goes as $1/g^2(\d u^*)^2$.
Considering a system at $r_s=5$, with $\Phi_2=-0.25$, we have $\d
u^*=3.279$. Assuming $g=5$ we have a probability of 0.0037 of the
system being in the strong-coupling phase.

\section{Appendix B}

Consider a generic theory with a fermion and a boson ($\psi$ and
$\sigmab$ in our problem), in which we integrate out the fermions
to get an effective action $S_{eff}$ for $\sigmab$.  It is clear
that this action gives all the answers for questions involving
$\sigmab$ alone. However, if at this stage we want to ask a
question involving the fermions, which have been integrated out,
how should one proceed? This is not an unprecedented situation,
but we discuss it nonetheless for completeness.

Consider the following schematic partition function in which the
fermions have been coupled to fermionic sources $J,{\bar J}$ to enable
a computation of their correlators: \beqr Z(J,\bar{J}) &=& \int
d\sigma d\psi d \bar{\psi} e^S\\ S(\sigma, \ \psi , \ \bar{\psi}) &=&
\bar{J}\psi + J \bar{\psi}+ \bar{\psi}(\p + M\sigma )\psi -\sigma^2 /2
\eeqr where $M$ is a coupling matrix, and $\p$ denotes the fermion
kinetic energy.  (In our case $(\p = i\omega -\ve_{\a} )$).

Upon integrating the fermions out we get \beq S_{eff}(\sigma ) =
\bar{J}{1 \over \p +M\sigma} J+ Tr \ln (\p +M \sigma )
-\sigma^2/2. \eeq

To obtain the fermion propagator $G$, we take the $J$ and $\bar{J}$
derivatives of $\ln Z$, set $J=\bar{J}=0$, and obtain \beq G =
(Z(0,0))^{-1}\int d\sigma
  {1 \over \p +M\sigma} e^{S_{eff}(\sigma )}.\eeq

Thus the full $G$ is just the propagator in a given external field
$\sigma$, functionally averaged of over $\sigma$ with its effective
action in the Boltzmann factor.

Let $S_{eff}$ have the following expansion about a saddle-point: \beq
  S_{eff}=S_{0}(\sigma_0) +{1\over 2} \delta \sigma D^{-1}(\sigma_0)
  \delta \sigma+\mbox{neglected terms}\eeq If we ignore the
  fluctuations we get \beq G_{SP} ={1 \over \p +M\sigma_0} \eeq If we
  keep the gaussian fluctuations, we get to lowest nontrivial order \beq
  G= {1 \over \p +M\sigma_0 }\left[ 1+{1\over 2} M {1 \over \p
  +M\sigma_0}M {1 \over \p +M\sigma_0}\langle \delta \sigma \ \delta
  \sigma
\rangle+\cdots \right]\eeq where the average  \beq \langle \delta
\sigma \ \delta \sigma \rangle = D(\sigma_0)\eeq is just the
$\sigma$ propagator in the gaussian approximation. Clearly, this
procedure generates all the Feynman diagrams containing the
three-point vertex coupling the fermions to fluctuations of
$\sigmab$. This series can be resummed in the usual way by
constructing a self-energy for the fermions, with the lowest
nontrivial contribution to it being Fig. 15. Thus there is no
double-counting involved in using Fig. 15 (or Eq.
(\ref{t0-Gamma})).

In the weak-coupling regime, $\sigma_0 =0$, and Eq. (\ref{dinv})
gives the inverse $\sigma$ propagator.  In the symmetry-broken
strong-coupling regime, the term $ \p +M\sigma_0 $ corresponds to
the hamiltonian of Eqn.  (\ref{hwithmeansigma}), the treatment of
which follows the equation.

 \end{document}